\documentclass{article}

\usepackage{amsmath}
\usepackage{amssymb}
\usepackage{amsthm}
\usepackage{amscd}
\usepackage{graphicx}

\newcommand{\hako}[1]{\fbox{\parbox[c][6.5pt][b]{4.5pt}{$ #1 $}}}

\newcommand{\g}{\mathfrak g}
\newcommand{\gh}{\hat{\g}}
\newcommand{\Z}{{\mathbb Z}}

\newtheorem{theorem}{Theorem}[section]
\newtheorem{lemma}[theorem]{Lemma}
\newtheorem{conjecture}[theorem]{Conjecture}
\newtheorem{proposition}[theorem]{Proposition}

\theoremstyle{definition}

\newtheorem{example}[theorem]{Example}

\numberwithin{equation}{section}

\makeatletter
\def\eqnarray{%
  \stepcounter{equation}%
  \let\@currentlabel=\theequation
  \global\@eqnswtrue
  \global\@eqcnt\z@
  \tabskip\@centering
  \let\\=\@eqncr
  $$\halign to \displaywidth\bgroup\@eqnsel\hskip\@centering
  $\displaystyle\tabskip\z@{##}$&\global\@eqcnt\@ne
  \hfil$\displaystyle{{}##{}}$\hfil
  &\global\@eqcnt\tw@$\displaystyle\tabskip\z@{##}$\hfil
  \tabskip\@centering&\llap{##}\tabskip\z@\cr}
\makeatother

\def\geh{\mathfrak{g}}

\def\cd{\cdots}
\def\ot{\otimes}

\def\veps{\varepsilon}
\def\vphi{\varphi}

\def\Z{{\mathbb Z}}
\def\et{\tilde{e}}
\def\ft{\tilde{f}}

\title{Soliton Cellular Automata Associated With Crystal Bases}

\author{
Goro Hatayama\thanks{
Institute of Physics, University of Tokyo, Komaba, Tokyo 153-8902, Japan},
Atsuo Kuniba,$\hspace{-1.2mm}^*$
and Taichiro Takagi\thanks{
Department of Mathematics and Physics, National Defense Academy,
Yokosuka 239-8686, Japan}
}

\date{}

\begin{document}

\maketitle

\begin{abstract}
We introduce a class of cellular automata associated with crystals of irreducible
finite dimensional representations of quantum affine algebras
$U'_q(\hat{\geh}_n)$.
They have solitons labeled by  crystals
of the smaller algebra $U'_q(\hat{\geh}_{n-1})$.
We prove stable propagation of one soliton for
$\hat{\geh}_n = A^{(2)}_{2n-1}, A^{(2)}_{2n},
B^{(1)}_n, C^{(1)}_n, D^{(1)}_n$ and $D^{(2)}_{n+1}$.
For $\gh_n = C^{(1)}_n$, we also prove
that the scattering matrices of two solitons
coincide with the combinatorial $R$ matrices
of $U'_q(C^{(1)}_{n-1})$-crystals.
\end{abstract}

\section{Introduction}\label{sec:intro}

Cellular automata are the dynamical systems in which
the dependent variables assigned to a space lattice take
discrete values  and evolve under a certain rule.
They exhibit rich behavior, which have been widely
investigated in physics, chemistry, biology and
computer sciences \cite{W}.
When the space lattice is one dimensional,
there are several examples known as the soliton cellular automata
\cite{FPS, PF, PAS, PST, TS, T}.
They possess analogous features to the solitons
in integrable non-linear partial differential equations.
For example,
some patterns propagate with fixed velocity
and they undergo collisions retaining their identity and
only changing their phases.

There is a notable progress recently
in understanding the integrable structure
in the soliton cellular automata.
In the papers \cite{TTMS,MSTTT,TNS} it was shown that
a class of soliton cellular
automata can be derived from the known soliton equations such as
Lotka-Volterra and Toda equations through a
limiting procedure called {\em ultra-discretization}.
The method enables one to construct the explicit solutions and the
conserved quantities
of the former from that of the latter.
A key in the ultra-discretization is the identities: $(a, b \in {\mathbb R})$
\begin{eqnarray*}
\lim_{\epsilon \rightarrow +0} \epsilon \log
(e^{\frac{a}{\epsilon}} + e^{\frac{b}{\epsilon}}) &=& \max(a,b), \\
\lim_{\epsilon \rightarrow +0} \epsilon \log
(e^{\frac{a}{\epsilon}} \times e^{\frac{b}{\epsilon}}) &=& a + b.
\end{eqnarray*}
In a sense they change $+$ into $\max$ and $\times$ into $+$.
This is a transformation of the continuous operations into piecewise linear ones
preserving the distributive law:
\begin{equation*}
(A+B)\times C = (A\times C) + (B \times C) \, \rightarrow \,
\max(a,b) + c = \max(a+c, b+c).
\end{equation*}
The non-uniqueness of the distributive structure
is noted by Sch\"utzenberger in combinatorics, where
the procedure corresponding to the inverse of the ultra-discretization is called
`tropical variable change' \cite{Ki}.

There is yet further intriguing aspect in the soliton cellular automata
(called `box and ball systems') in  \cite{T, TNS}.
There the scattering of two solitons is described by the rule which turns out to be
identical with the $U_q'(A^{(1)}_n)$ combinatorial $R$ matrix
\cite{NY} from the crystal base theory.
The latter has an origin in the quantum affine algebras at $q=0$,
where the representation theory is piecewise linear in a certain sense.

Motivated by these observations we formulate in this paper
and \cite{HHIKTT} a class of
cellular automata directly in terms of crystals and link the subject to
the $1+1$ dimensional quantum integrable systems.
The theory of crystals is invented by Kashiwara \cite{Kas} as a
representation theory of the quantized Kac-Moody algebras at $q = 0$.
It is a powerful tool that reduces many essential problems
into combinatorial questions on the associated crystals.
Irreducible decomposition of tensor products and the
Robinson-Schensted-Knuth correspondence are typical such
problems \cite{DJM,KN,N}.
By connecting the classical and affine crystals,
it also explains \cite{KMN1, KMN2}
the appearance of the affine Lie algebra characters \cite{DJKMO}
in  Baxter's corner transfer matrix method in solvable lattice models \cite{B}.

Here we shall introduce a cellular automaton associated with
crystals of irreducible finite dimensional representations of
quantum affine algebras $U'_q(\hat{\geh}_n)$.
The basic idea is to regard the time evolution
in the automaton as the action of a row-to-row transfer matrix of
integrable $U'_q(\hat{\geh}_n)$ vertex models at $q=0$.
The essential point is to consider the tensor product of crystals {\em not} around
the `anti-ferromagnetic vacuum' as in \cite{KMN1,KMN2}, but
rather in the vicinity of the `ferromagnetic vacuum'.

Let $B$ be a classical crystal of
irreducible finite dimensional representations of the
quantum affine algebra $U'_q(\hat{\geh}_n)$.
It is a  finite set having a weight decomposition and
equipped with
the maps $\et_i, \ft_i: B \rightarrow B \sqcup \{0\}$ and
$\veps_i, \vphi_i: B \rightarrow {\mathbb Z}_{\ge 0}$
$(i \in \{0,1,\ldots, n\})$ satisfying certain axioms.
(cf. Definition 2.1 in \cite{KKM}.)
For two crystals $B$ and $B'$ the tensor products $B' \otimes B$ and $B \otimes B'$
are again crystals which are canonically isomorphic.
The isomorphism $B' \otimes B
\stackrel{\sim}{\rightarrow} B\otimes B'$ is called the
combinatorial $R$ matrix
%%%%%%%%%%%%%%%%%%%%%%%%%%%%%%%%%%%%%
\footnote{More precisely, the isomorphism combined with the data on the
energy function is called the combinatorial $R$ matrix \cite{KMN1}.}.
%%%%%%%%%%%%%%%%%%%%%%%%%%%%%%%%%%%%%
%
Suppose that there are special elements
denoted by $1 \in B$ and $u_\natural \in B'$ with the properties
\begin{eqnarray*}
&({\mathrm I})& \quad u_\natural \otimes 1 \simeq 1 \otimes u_\natural,\\
&(\mathrm{II})& \quad \text{For any } u \in B'\,  \text{ there exists } \,
k \in {\mathbb Z}_{\ge 0} \text{ such that } \\
&&\quad u \otimes \overbrace{1 \otimes \cdots \otimes 1}^{k}
\simeq b_1 \otimes \cdots \ot b_{k}\otimes u_\natural,\qquad
  (b_i \in B).
\end{eqnarray*}
We take dynamical variables of our automaton  from the crystal $B$
and regard their array
$\ldots, b^t_{-1}, b^t_0, b^t_1, \ldots$ at time $t$  as an element
of the tensor product of crystals
\begin{equation*}
\cd \ot b^t_{-1} \ot  b^t_0 \ot  b^t_1 \ot \cd  \in
\cdots \otimes B \otimes B \otimes B \otimes \cdots,
\end{equation*}
where we assume the boundary condition
$b^t_j = 1 \in B$ for $\vert j \vert \gg 1$.
See Section \ref{subsec:cellauto} for a precise treatment.
The `ferromagnetic' state $\forall b^t_i = 1$ is understood as
the vacuum of the automaton.
The infinite tensor product with such a boundary condition does not admit a
crystal structure.
Nevertheless one can make sense of the construction below
thanks to the properties (I) and (II).
The time evolution is induced by sending
$u_\natural$ from  left to  right
via the repeated application of the combinatorial $R$ matrix as
\begin{alignat}{22}
B' & \ot & (\cd & \ot &B& \ot &B& \ot &B& \ot & \cd) & \simeq &
(\cd & \ot &B& \ot &B& \ot &B& \ot & \cd) & \ot & B' \nonumber\\
u_\natural & \ot & (\cd & \ot &b^t_{-1}& \ot &b^t_0& \ot &b^t_1& \ot & \cd)
& \simeq &
(\cd & \ot &b^{t+1}_{-1}& \ot &b^{t+1}_0& \ot &b^{t+1}_1& \ot & \cd)
& \ot & u_\natural, \nonumber
\end{alignat}
which is well-defined as long as the above properties and the boundary
conditions are fulfilled.
In the language of the quantum inverse scattering method \cite{STF,KS},
this is the action of  the
$q=0$ row-to-row transfer matrix whose
auxiliary and quantum spaces are labeled by $B'$ and
$\cd \ot B \ot B \ot \cd$, respectively.
Note that the transfer matrix has
effectively reduced to the  $(u_\natural, u_\natural)$-component of the
monodromy matrix since its action is considered under the
ferromagnetic boundary condition.
The fundamental case $\gh_n = A^{(1)}_n$ will be studied
in a more general setting in \cite{HHIKTT}.
In this paper we concentrate on the other non-exceptional series
\begin{align*}
\hat{\geh}_n &= A^{(2)}_{2n-1}, A^{(2)}_{2n},
B^{(1)}_n, C^{(1)}_n, D^{(1)}_n, D^{(2)}_{n+1},\\
\intertext{with the following choice of crystals:}
B &= B_1 \ni 1, \quad B' = B_\natural \ni u_\natural.
\end{align*}
Here $B_1$ is the crystal associated with the vector
representation of the classical subalgebra of
$\hat{\geh}_n$ except for $A^{(2)}_{2n}$ and $D^{(2)}_{n+1}$.
Their cardinalities are $\sharp B = 2n, 2n+1, 2n+1, 2n, 2n$ and $2n+2$,
respectively.
The element $1 \in B_1$ is the highest weight one
%%%%%%%%%%%%%%%%%%%%%%%%%%%%%%%%%%%%%%%
\footnote{For $(\gh_n, B_1)$ treated in this paper,
a parallel construction seems possible also with the choice
$1 =$ lowest weight element.}.
%%%%%%%%%%%%%%%%%%%%%%%%%%%%%%%%%%%%%%%
%
To explain $B_\natural$ and $u_\natural$, recall the coherent family
$\{B_l \mid l \in {\mathbb Z}_{\ge 1}\}$ of the
perfect crystals obtained in \cite{KKM}.
It contains the $B_1$ as its first member
%%%%%%%%%%%%%%%%%%%%%%%%%%%%%%%%%%%%%
\footnote{For $C^{(1)}_n$, the family in \cite{KKM}
does not contain $B_1$. See \cite{HKKOT}.}.
%%%%%%%%%%%%%%%%%%%%%%%%%%%%%%%%%%%%%%
%
The $B_l$ with higher $l$ corresponds to an
$l$-fold symmetric fusion of $B_1$.
Then $B_\natural$ in question is an infinite set corresponding to a certain
$l \rightarrow \infty$ limit of
$B_l$ and $u_\natural$ is its highest weight element
%%%%%%%%%%%%%%%%%%%%%%%%%%%%%%%%%%%%%%%%%%%%
\footnote{They are different from
the limits $B_\infty$ and $b_\infty$ in  \cite{KKM}.}.
%%%%%%%%%%%%%%%%%%%%%%%%%%%%%%%%%%%%%%%%%%%%%.
%
We shall call the resulting dynamical system
$U'_q(\hat{\geh}_n)$ automaton.
They are essentially solvable trigonometric vertex models
at $q = 0$ in the vicinity of the ferromagnetic vacuum.
A peculiarity here is the
extreme anisotropy with respect to the relevant fusion degrees;
$B_1$ is the simplest one, while $B' = B_\natural$
corresponds to an infinite fusion
%%%%%%%%%%%%%%%%%%%%%%%%%%%%%%%%%%%%%%%%%%%%%
\footnote{To take  $B' =  B_l$ with finite $l$ is an interesting
generalization. See  \cite{HHIKTT} for $A^{(1)}_n$ case.}.
%%%%%%%%%%%%%%%%%%%%%%%%%%%%%%%%%%%%%%%%%%%%%

Once the automata are constructed
the first question will be if they are solitonic.
We prove a theorem that
\begin{itemize}
\item the $U'_q(\hat{\geh}_n)$ automaton has the patterns
labeled by the crystals
$\{ B_l\}$ of the algebra $U'_q(\hat{\geh}_{n-1})$ that
propagate stably with velocity $l$.
\end{itemize}
Computer experiments indicate that they indeed behave like solitons.
For instance, the initially separated patterns labeled by the
$U'_q(\gh_{n-1})$-crystal elements $b \in B_l$ and $c \in B_k$ $(l > k)$
undergo a scattering into two patterns labeled again by some
$c' \in B_k$ and $b' \in B_l$.
Let ${\mathcal S}: B_l\ot B_k \rightarrow B_k \ot B_l$
 be the two-body scattering matrix of such collisions,
namely, ${\mathcal S}(b \ot c) = c'\ot b'$.
Let ${\mathcal R}: B_l\ot B_k \rightarrow B_k \ot B_l$ denote the
combinatorial $R$ matrix of $U'_q(\gh_{n-1})$.
Then we prove
\begin{itemize}
\item ${\mathcal S} = {\mathcal R}$
\end{itemize}
for $\gh_n = C^{(1)}_n$ and conjecture it for all the  other $\gh_n$.
Similarly the scattering of multi-solitons
labeled by $B_{l_1}, \ldots, B_{l_N}$ ($l_1 > \cd > l_N$)
is  given by the isomorphism
$B_{l_1} \ot \cd \ot B_{l_N} \simeq B_{l_N} \ot \cd \ot B_{l_1}$
experimentally.
Thus the solitonic nature is guaranteed by the Yang-Baxter
equation obeyed by ${\mathcal S} = {\mathcal R}$.
A precise formulation of these claims is done through an injection
\begin{equation*}
\imath_l : U'_q(\gh_{n-1})\text{-crystal }B_{l} \rightarrow
(U'_q(\gh_n)\text{-crystal }B_{1})^{\otimes l},
\end{equation*}
which will be described in Section \ref{sec:conj}.

Admitting that they are soliton cellular automata,
the second question is if there exist classical integrable equations
governing them, possibly via the ultra-discretization.
Here we only confirm this for
$A^{(2)}_2$ case by relating the associated automaton to
the known $A^{(1)}_1$ example \cite{TS}.
This observation is due to \cite{HI}.

The layout of the paper is as follows.
In Section \ref{sec:a22} we first explain our construction of
the $U'_q(\hat{\geh}_n)$ automata concretely
along the $\hat{\geh}_n = A^{(2)}_2$ example.
It is valid for any $U'_q(\hat{\geh}_n)$ and any
finite crystals having the properties (I) and (II).
In Section \ref{sec:conj}, we formulate the theorem and
the conjecture for $\hat{\geh}_n$ precisely.
We sketch a proof of ${\mathcal S} = {\mathcal R}$ for $C^{(1)}_n$ case.
In principle the idea used in the proof can also be used for the
other $\gh_n$.
We will specify $B_\natural$  as an infinite set  with
the actions $\et_i, \ft_i: B_\natural \rightarrow B_\natural \sqcup \{0\}$ but
without the maps
$\veps_i, \vphi_i$.
In Section \ref{sec:dis} concluding remarks are given.
Appendix \ref{app:natural} is devoted to an explanation of
what is meant by `$B_\natural \ot B_1 \simeq B_1 \ot B_\natural$', which shows up
when the infinite set $B_\natural$ is substituted into the finite crystal $B'$.
This is actually abuse of notation meaning an invertible map
$R': B_\natural \times B_1 \rightarrow  B_1 \times B_\natural$ between the sets.
We state a conjecture
on a stability of the combinatorial $R$ matrix
$B_l \ot B_1 \simeq B_1 \ot B_l$ when $l$ gets large,
which ensures the well-definedness of the map $R'$.
It assures that we may regard $B_\natural$ as
a finite crystal $B_l$ with a sufficiently large $l$ to define our
automata.

Our construction here and in \cite{HHIKTT} is a crystal interpretation of the
$L$-operator approach \cite{HIK}
for a $\hat{\geh}_n = A^{(1)}_n$ case.
The $U'_q(A^{(1)}_n)$ automaton in this sense
coincides with the ones in \cite{TS,T,TNS}.
As in the $C^{(1)}_n$ case in this paper,
the properties stated in the above can actually be proved
by means of the crystal theory.
The detail will appear elsewhere along with the results on a more general choice of
the crystals $B$ and $B'$ \cite{HHIKTT}.

\vspace{0.4cm}
\noindent
{\bf Acknowledgements} \hspace{0.1cm}
The authors thank K. Hikami, R. Inoue, T. Nakashima, M. Okado,
T. Takebe, T. Tokihiro, Z. Tsuboi and Y. Yamada for valuable discussions.

\section{ $A^{(2)}_2$ example}\label{sec:a22}

Let us explain our automata concretely along the  case
$\gh_n = A^{(2)}_2$.
This simple example is helpful to gain the idea for the general
$\gh_n$ case treated in the next section.

As a peculiarity in the rank 1 situation, the $U'_q(A^{(2)}_2)$
automaton turns out to be an `even time sector' of  \cite{TS}.

\subsection{$A^{(2)}_2$ crystals}\label{subsec:a22crystal}

\begin{figure}
\includegraphics[width=10cm]{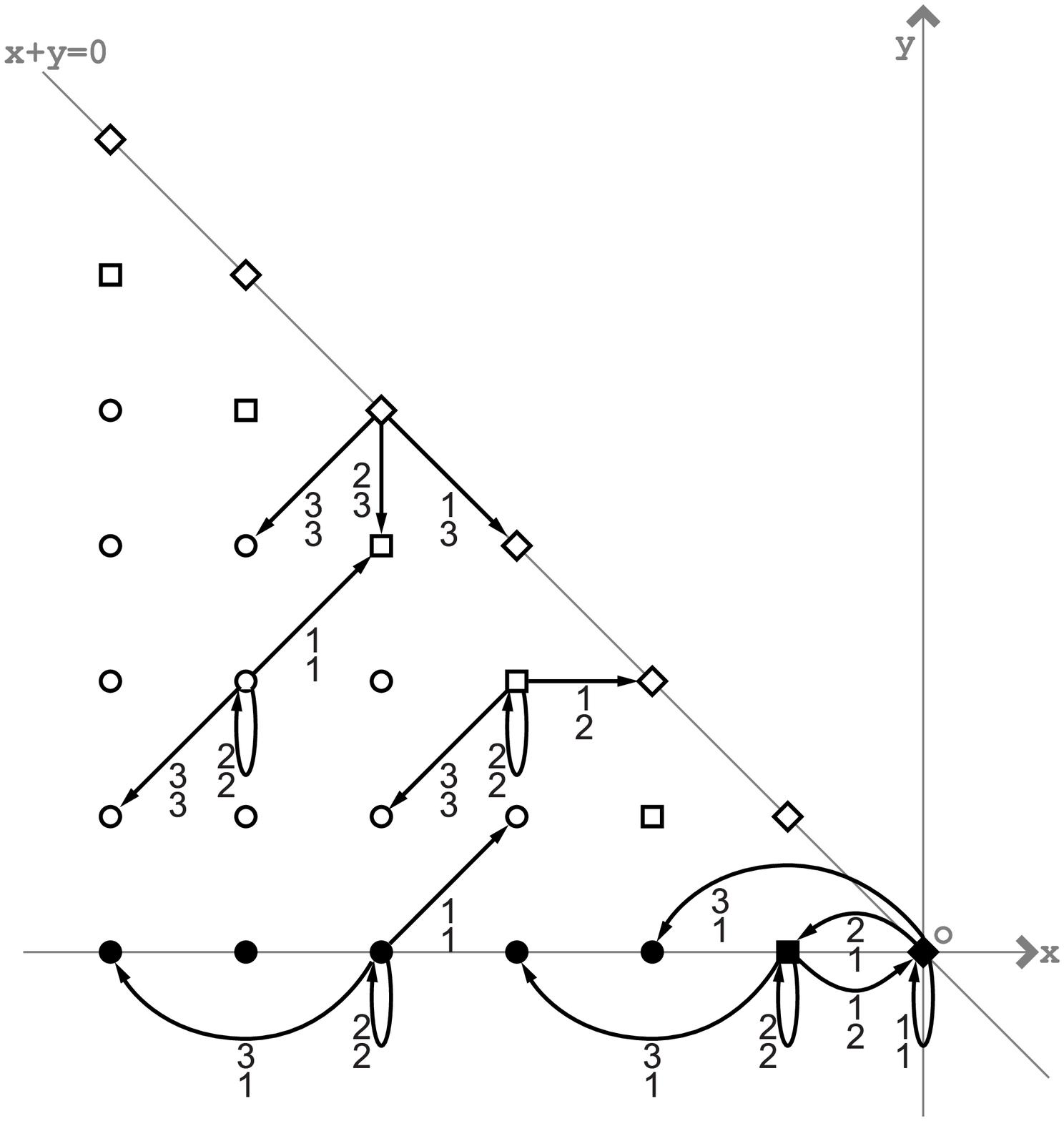}
\caption{}{The Semi-infinite triangle representing $R'$. There are 6 different patterns depicted by circles, squares and diagonal squares which are filled or empty.}
 \label{fig:sit}
\end{figure}

For $l \in {\mathbb Z}_{\ge 1}$ set
\begin{equation}\label{eq:bl}
B_l = \{ (x,y) \mid x, y \in {\mathbb Z}_{\ge 0},\, x + y \le l \}.
\end{equation}
The action of the Kashiwara operators
${\tilde e}_i, {\tilde f}_i (i=0,1) : B_l \rightarrow B_l \sqcup \{0\}$
are given by
\begin{eqnarray*}
{\tilde e}_0 (x,y) & = & \begin{cases}
 (x-1,y) & \text{if $x > y$},\\
 (x,y+1) & \text{if $x \le y$},
 \end{cases}\quad
{\tilde e}_1 (x,y)  =  (x+1,y-1),\\
{\tilde f}_0 (x,y) & = & \begin{cases}
 (x+1,y) & \text{if $x  \ge y$},\\
 (x,y-1) & \text{if $x < y$},
 \end{cases}\quad
{\tilde f}_1 (x,y)  =  (x-1,y+1).\\
\end{eqnarray*}
In the above, the right hand sides are to be understood as $0$
if they are not in $B_l$.
Setting $\veps_i(b) = \max_k\{\et_i^k b \neq 0 \mid k \ge 0 \}$ and
$\vphi_i(b) = \max_k\{\ft_i^k b \neq 0 \mid k \ge 0 \}$, one has
\begin{alignat*}{2}
\veps_0(b) &= l-x-y+2(x-y)_+, \quad \veps_1(b) &= y,\\
\vphi_0(b) &= l-x-y+2(y-x)_+, \quad \vphi_1(b) &= x,
\end{alignat*}
for $b = (x,y) \in B_l$.
Here the symbol $(\cdot)_+$ stands for
\begin{equation*}
(x)_+ = \max(0,x).
\end{equation*}
These results are obtained by extrapolating the $A^{(2)}_{2n}$ result
\cite{KKM} to $n=1$.
For $l=1$ we use a simpler notation as
\begin{equation}\label{eq:simpler}
B = B_1, \quad 1 = (1,0), \, 2 = (0,0), \, 3 = (0,1).
\end{equation}
Given two crystals $B$ and $B'$, one can form another crystal (tensor product)
$B \ot B'$ \cite{Kas}.
The crystal $B_l \ot B_1$ is connected and so is  $B_1 \ot B_l$.
Calculating the map $B_l \ot B_1 \rightarrow B_1 \ot B_l$ commuting with
$\et_i$ and $\ft_i$, one has
\begin{proposition}\label{pr:a22combinatorialR}
The combinatorial $R$ matrix
$B_l \otimes B \simeq B \otimes B_l$ is given by
\begin{eqnarray*}\label{eq:isol1}
(x,y)\otimes 1 &\simeq& \begin{cases}
1 \otimes (l,0) & \text{if $(x,y) = (l,0)$},\\
3 \otimes (x+1,y-1) & \text{if $x+y=l, \,y \ge 1$}, \\
2 \otimes (x+1,y) & \text{if $x + y = l-1$},\\
1 \otimes (x+1,y+1) &\text{otherwise},
\end{cases}\\
(x,y) \otimes 2 &\simeq& \begin{cases}
1 \otimes (l-1,0) &\text{if $(x,y) = (l,0)$},\\
3 \otimes (x,y-1) &\text{if $x+y = l,\,  y \ge 1$},\\
2 \otimes (x,y) &\text{otherwise},
\end{cases}\\
(x,y) \otimes 3 &\simeq& \begin{cases}
3 \otimes (0,l) &\text{if $(x,y) = (0,l)$},\\
2 \otimes (0,l) &\text{if $(x,y) = (0,l-1)$},\\
1 \otimes (0,1) &\text{if $(x,y) = (1,0)$},\\
1 \otimes (x-2,y) &\text{if $2 \le x \le l, \, y=0$},\\
1 \otimes (x, y+2) &\text{if $x=0, \, 0 \le y \le l-2$},\\
3 \otimes (x-1,y-1) &\text{otherwise}.
\end{cases}
\end{eqnarray*}
\end{proposition}
In Section \ref{subsec:cellauto}
we shall use formal $l \rightarrow \infty$ limits
of $B_l$ and the combinatorial $R$ matrix
$B_l \otimes B \simeq B \otimes B_l$.
In the present case
the prescription is to simply shift the coordinate $(x,y)$ to $(x-l,y)$
and to consider
$$B_\natural = \{ (x,y) \mid x
\in {\mathbb Z}_{\le 0}, y \in {\mathbb Z}_{\ge 0}, x + y \le 0 \},
$$
without specifying a crystal structure.
The map $R': B_\natural \otimes B \simeq B \otimes B_\natural$
in the sense of Appendix \ref{app:natural}
is deduced from Proposition \ref{pr:a22combinatorialR}
by concentrating on those $(x,y)$ in the vicinity of $(0,0)$.
Thus it reads
\begin{eqnarray}
(x,y)\otimes 1 &\simeq& \begin{cases}
1 \otimes (0,0) & \text{if $(x,y) = (0,0)$},\\
3 \otimes (x+1,y-1) & \text{if $x+y=0, \,y \ge 1$}, \\
2 \otimes (x+1,y) & \text{if $x + y = -1$},\\
1 \otimes (x+1,y+1) &\text{otherwise},
\end{cases}\label{eq:isoinfty1}\\
(x,y) \otimes 2 &\simeq& \begin{cases}
1 \otimes (-1,0) &\text{if $(x,y) = (0,0)$},\\
3 \otimes (x,y-1) &\text{if $x+y = 0,\,  y \ge 1$},\\
2 \otimes (x,y) &\text{otherwise},
\end{cases}\label{eq:isoinfty2}\\
(x,y) \otimes 3 &\simeq& \begin{cases}
1 \otimes (x-2,y) &\text{if $y=0$},\\
3 \otimes (x-1,y-1) &\text{otherwise}.
\end{cases}\label{eq:isoinfty3}
\end{eqnarray}
To depict this in a figure we put
${b \atop b'}$ beside the arrow
$(x,y) \rightarrow (x',y')$ to signify the
relation
\begin{equation*}
R': (x,y) \ot b \xrightarrow{\sim} b' \ot (x',y').
\end{equation*}
We call $b$ and $b'$ the upper index and the lower index, respectively.
Now (\ref{eq:isoinfty1})--(\ref{eq:isoinfty3}) are summarized in
the semi-infinite triangle in Figure \ref{fig:sit}.

\subsection{Cellular automaton} \label{subsec:cellauto}

By applying $B_\natural \ot B \simeq B \ot B_\natural$ successively
one has
\begin{alignat}{6}
B_\natural & \ot &  \overbrace{B \ot  \cd \ot B}^{L} & \simeq &
\overbrace{B \ot \cd  \ot B}^{L} & \ot & B_\natural \nonumber\\
u  & \ot & b_{i} \ot  \cd \ot b_{j}
& \simeq  &
b'_{i} \ot \cd \ot b'_{j} & \ot&u'
\label{eq:isomulti}
\end{alignat}
for any $L \in {\mathbb Z}_{\ge 1}$, where
$i=-[\frac{L}{2}], j = i+L-1$
%%%%%%%%%%%%%%%%%%%%%%%%%%%%%%%%%%%%%%%%%%
\footnote{The symbol $[x]$ denotes the largest integer not
exceeding $x$.}.
%%%%%%%%%%%%%%%%%%%%%%%%%%%%%%%%%%%%%%%%%%
%
Denote the element $(0,0) \in B_\natural$ by $u_\natural$.
The map  $R': B_\natural \otimes B \simeq B \otimes B_\natural$
(\ref{eq:isoinfty1})--(\ref{eq:isoinfty3}) has the properties (I) and (II)
in Section \ref{sec:intro}.
Set
\begin{equation*}
{\mathcal P} = \{ b: {\mathbb Z} \rightarrow B \mid
b_j = 1 \in B \text{ for } \vert j \vert \gg 1 \}.
\end{equation*}
We shall regard ${\mathcal P}$ as a
subset of the tensor product which is formally infinite in both directions, i.e.,
\begin{equation}\label{eq:defp}
{\mathcal P} = \{ \cdots \ot b_{-1}\otimes b_0 \otimes b_1 \otimes \cdots
\in \cdots \ot B \otimes B \otimes B \otimes \cdots \mid
b_j = 1 \text{ for } \vert j \vert \gg 1 \}.
\end{equation}
In the latter picture one should distinguish the elements
even though they are the same under translations.
For example,
\begin{eqnarray*}
&&\cd 1 \ot 1 \ot 1 \ot 2 \ot 3 \ot 1 \ot 1 \ot 1 \ot \cd,\\
&&\cd 1 \ot 1 \ot 1 \ot 1 \ot 2 \ot 3 \ot 1 \ot 1 \ot \cd
\end{eqnarray*}
are distinct elements in ${\mathcal P}$.
The set ${\mathcal P}$ (\ref{eq:defp}) is not equipped with a
crystal structure.
Nevertheless the properties (I) and (II) enable us to define an invertible map
$T: {\mathcal P} \rightarrow {\mathcal P}$ that formally corresponds
to an $L \rightarrow \infty$ limit of (\ref{eq:isomulti}).
To describe it precisely, note that any element in ${\mathcal P}$
has the form
\begin{equation}\label{eq:p}
p = \cd \ot 1 \ot 1 \ot b_i \ot \cd \ot b_j \ot 1 \ot 1 \ot \cd
\quad (i \le j),
\end{equation}
where $b_i, b_{i+1}, \ldots, b_j \in B$.
Owing to the properties (I) and (II) there exists $k_0 \in {\mathbb Z}_{\ge 0}$
such that
\begin{equation*}
u_\natural \ot b_i \ot \cd \ot b_j \ot \overbrace{1 \ot \cd \ot 1}^k
\simeq b'_i \ot \cd \ot b'_j \ot b'_{j+1} \ot \cd \ot b'_{j+k} \ot u_\natural
\end{equation*}
for all $k \ge k_0$.
Then $T(p) \in {\mathcal P}$ is defined by
\begin{equation*}
T(p) = \cd \ot 1 \ot 1 \ot b'_i \ot \cd \ot b'_j \ot b'_{j+1} \ot \cd \ot b'_{j+k}
 \ot 1 \ot 1 \ot \cd,
\end{equation*}
which is $k$-independent as long as $k \ge k_0$.
The inverse $T^{-1}$ can be described similarly.

The map $T$  plays the role of the `time evolution' operator.
It is a $q=0$  analogue of the
row-to-row transfer matrix of a solvable lattice model
in the vicinity of the ferromagnetic vacuum.

Given $p \in {\mathcal P}$ in (\ref{eq:p}) define $u_m \in B_\natural$
for all $m \in {\mathbb Z}$ by the recursion relation and the
boundary condition
\begin{eqnarray*}
u_{m-1} \ot b_m & \simeq & b'_m \ot u_m \quad \text{ for all } \, m \in {\mathbb Z},\\
u_m & = & u_\natural \quad \qquad \text{ for }\, m \le i-1.
\end{eqnarray*}
Plainly,
$u_\natural \ot (\cd \ot b_{m-1}\ot b_m \ot b_{m+1} \ot \cd)
\simeq \cd \ot b'_{m-1} \ot b'_m \ot u_m \ot b_{m+1} \ot \cd$.
Due to the properties (I) and (II) the sequence
$u_m, u_{m+1}, \ldots$ tends to $u_\natural = (0,0) \in B_\natural$.
In this way any element $p \in {\mathcal P}$ specifies
a {\em trajectory} $\{u_m\}_{m=-\infty}^\infty$ in the
semi-infinite triangle (Figure \ref{fig:sit})
that starts at the origin $(0,0)$ and returns to it finally.
This picture is useful in calculating $T(p)$.
Namely, the trajectory is determined by following the arrows with
the upper indices $\ldots, b_{m-1}, b_m, b_{m+1}, \ldots$ appearing
in $p = \cd \ot b_{m-1}\ot b_m \ot b_{m+1} \ot \cd$.
Then $T(p)$ is constructed by tracing their lower indices as
$T(p) = \cd \ot b'_{m-1} \ot b'_m \ot b'_{m+1} \ot \cd$.

For $p = \cd \ot b_j \ot b_{j+1} \ot \cd \in {\mathcal P}$
and $t \in {\mathbb Z}$
define $b^t_j \in B$ by $T^t(p) = \cd \ot b^t_j \ot b^t_{j+1} \ot \cd$.
Then the time evolution of the cellular automaton is displayed with  the arrays
\begin{equation*}
\begin{array}{ccccccc}
%&&& \ldots &&&\\
\ldots & b^0_{-2} & b^0_{-1} & b^0_0 & b^0_1 & b^0_2 &\ldots \\
\ldots & b^1_{-2} & b^1_{-1} & b^1_0 & b^1_1 & b^1_2 &\ldots \\
\ldots & b^2_{-2} & b^2_{-1} & b^2_0 & b^2_1 & b^2_2 &\ldots \\
%&&& \ldots.&&&
\end{array}
\end{equation*}
Let us present a few examples.

\begin{example}\hfill
\begin{center}
t=0 : $\cdots$11233311111111111111111111111$\cdots$\\
t=1 : $\cdots$11111111123331111111111111111$\cdots$\\
t=2 : $\cdots$11111111111111112333111111111$\cdots$\\
t=3 : $\cdots$11111111111111111111111233311$\cdots$\\
\end{center}
\end{example}

\begin{example}\hfill
\begin{center}
t=0 : $\cdots$11333311111111111111111111111111$\cdots$\\
t=1 : $\cdots$11111111113333111111111111111111$\cdots$\\
t=2 : $\cdots$11111111111111111133331111111111$\cdots$\\
t=3 : $\cdots$11111111111111111111111111333311$\cdots$\\
\end{center}
\end{example}

\begin{example}\hfill
\begin{center}
t=0 : $\cdots$1133311111123111111111111111111111$\cdots$\\
t=1 : $\cdots$1111111133311123111111111111111111$\cdots$\\
t=2 : $\cdots$1111111111111133123311111111111111$\cdots$\\
t=3 : $\cdots$1111111111111111123111133311111111$\cdots$\\
t=4 : $\cdots$1111111111111111111123111111133311$\cdots$\\
\end{center}
\end{example}

\begin{example}\label{3SolitonExample}\hfill
\begin{center}
t=0 : $\cdots$11233111331111111111111211111111111111111111111$\cdots$\\
t=1 : $\cdots$11111112331133111111111121111111111111111111111$\cdots$\\
t=2 : $\cdots$11111111111133112331111112111111111111111111111$\cdots$\\
t=3 : $\cdots$11111111111111113311123311211111111111111111111$\cdots$\\
t=4 : $\cdots$11111111111111111111331111223311111111111111111$\cdots$\\
t=5 : $\cdots$11111111111111111111111133121111233111111111111$\cdots$\\
t=6 : $\cdots$11111111111111111111111111121331111112331111111$\cdots$\\
t=7 : $\cdots$11111111111111111111111111112111133111111123311$\cdots$\\
\end{center}
\end{example}

\begin{example}\label{3SolitonExample2}\hfill
\begin{center}
t=0 : $\cdots$11233111111133111121111111111111111111111111111$\cdots$\\
t=1 : $\cdots$11111112331111113312111111111111111111111111111$\cdots$\\
t=2 : $\cdots$11111111111123311112133111111111111111111111111$\cdots$\\
t=3 : $\cdots$11111111111111111233211113311111111111111111111$\cdots$\\
t=4 : $\cdots$11111111111111111111211233111331111111111111111$\cdots$\\
t=5 : $\cdots$11111111111111111111121111112331133111111111111$\cdots$\\
t=6 : $\cdots$11111111111111111111112111111111133112331111111$\cdots$\\
t=7 : $\cdots$11111111111111111111111211111111111113311123311$\cdots$\\
\end{center}
\end{example}
The last two show the independence of the order of collisions.
These examples suggest that the following patterns are stable
($Q \in {\mathbb Z}_{\ge 1}, R = 0,1$):
\begin{alignat*}{3}
\cdots \otimes & \overbrace{2 \otimes 3 \otimes \cdots \otimes 3}^{Q}
&\otimes \cdots & \quad R = 1,
\\
\cdots \otimes & \overbrace{3 \otimes 3 \ot  \cdots \otimes 3}^{Q}
& \otimes \cdots
&\quad R = 0.
\end{alignat*}
The both patterns should not be followed by $3$.
The former pattern  can be preceded by any element in $B$  while
the latter  should only be preceded by $1$.
$Q$ is the size of the soliton and $R$ is the number of
occurrences of $2$
in its front.
They move to the right with the velocity $2Q-R$ when separated
sufficiently.
These features are consistent with  $\gh_n = A^{(2)}_{2}$ case of
Theorem \ref{th:one}.
See also  Section \ref{subsec:data}.

In fact the $U'_q(A^{(2)}_2)$ automaton described above
can be interpreted \cite{HI} as an `even time sector' of the
automaton in \cite{TS}.
Replace the array of $\{1,2,3\}$ by that of $\{1, 2 \}$
with double length via the rule
$1 \rightarrow 11,
2 \rightarrow 12$ and
$3 \rightarrow 22$.
In the resulting array, play the `box and ball game' as in \cite{TS}.
Namely, we regard the array as a sequence of
cells which contains a ball or not according to
the array variable is $2$ or $1$, respectively.
In each time step, we move each  ball once
to the nearest right empty box starting from the leftmost ball.
Then the 2 time steps in the box and ball system
yield the 1 time step in our $U'_q(A^{(2)}_2)$ automaton.

In terms of crystals, this can be explained as follows.
First, the box and ball game in \cite{TS} is known \cite{HHIKTT}
to be equivalent to the $U'_q(A^{(1)}_1)$ automaton.
Let
\begin{equation*}
\hat{B}_k = \{ m_1\ldots m_k \mid m_i \in \{1, 2\},
m_1 \le \cd \le m_k \}
\end{equation*}
denote the $U'_q(A^{(1)}_1)$-crystal corresponding to the
$k$-fold symmetric tensor representation.
(We have omitted the frame of the usual semistandard tableaux.)
Consider the maps $h_\natural$ and $h_1$ defined by
\begin{align*}
h_\natural: &\quad B_\natural \quad \longrightarrow \quad\qquad\quad\;
\hat{B}_k \ot \hat{B}_k \quad (k \gg 1)\\
&(x,y) \quad \mapsto \quad \overbrace{1\ldots 1}^{k-2y}\overbrace{2\ldots 2}^{2y} \ot
\overbrace{1\ldots 1}^{k+x+y}\overbrace{2\ldots 2}^{-x-y},\\
h_1: & \quad B_1 \longrightarrow \hat{B}_1 \ot \hat{B}_1\\
&\quad 1 \quad \mapsto \quad 1 \ot 1\\
&\quad 2 \quad \mapsto \quad 1 \ot 2\\
&\quad 3 \quad \mapsto \quad 2 \ot 2.
\end{align*}
Then for $k$ large enough, we have the commutative diagram:
\begin{equation*}
\begin{CD}
B_\natural \ot B_1 @>{h_\natural\ot h_1}>> \hat{B}_k \ot \hat{B}_k
\ot \hat{B}_1 \ot \hat{B}_1
\\
@V{R'}VV  @VVV \\
B_1 \ot B_\natural @>{h_1 \ot h_\natural}>>
\hat{B}_1 \ot \hat{B}_1 \ot \hat{B}_k \ot \hat{B}_k,
\end{CD}
\end{equation*}
where the down arrow in the right column is the crystal isomorphism.
This asserts that the square of the $T$ in the
$U'_q(A^{(1)}_1)$ automaton coincides with the $T$ of  the
$U'_q(A^{(2)}_2)$ automaton.

%%%%%%%%%%%%%%%%%%%%%%%%%%%%%%%%%%%%%%%%%%%%
\section{$U'_q(\gh_n)$ automaton}\label{sec:conj}

\subsection{Theorem and conjecture}
\label{subsec:cellautoaffine}

Let us proceed to the $U'_q(\hat{\geh}_n)$ automata for
$\gh_n = A^{(2)}_{2n-1}\; (n \ge 3), A^{(2)}_{2n}\; (n \ge 1),
B^{(1)}_n \;(n  \ge 3), C^{(1)}_n\; (n \ge 2), D^{(1)}_n\; (n \ge 4)$
and $D^{(2)}_{n+1}\; (n \ge 2)$.
Our aim here is to formulate
the theorem and the conjecture stated in Section \ref{sec:intro} precisely.
In principle construction of the automata is the
same as the  $A^{(2)}_2$ case
explained in Section \ref{subsec:cellauto}.
The time evolution operator $T$ is constructed from the invertible map
$R': B_\natural\ot B_1 \rightarrow B_1 \ot B_\natural$
in the sense of Appendix \ref{app:natural}.
However its analytic form like (\ref{eq:isoinfty1})--(\ref{eq:isoinfty3})
is yet unknown for general $\gh_n$
%%%%%%%%%%%%%%%%%%%%%%%%%%%%%%%%%%%%%%%%%%%%%
\footnote{For $C^{(1)}_n$ we have a concrete description of the
combinatorial $R$ matrix $B_l \ot B_k \simeq B_k \ot B_l$
in terms of an insertion algorithm.
See \cite{HKOT}.}.
%%%%%%%%%%%%%%%%%%%%%%%%%%%%%%%%%%%%%%%%%%%%
%
Thus we have generated the combinatorial
$R$ matrix $R: B_l \ot B_1 \simeq B_1 \ot B_l$ directly by computer,
and investigated the automata associated with $R$ instead of $R'$
for several large $l$.
Consistently with Conjecture \ref{con:atability},
their behaviour becomes stable when $l$ gets large,
yielding our automata associated with $R'$.

What we present in Section \ref{subsec:data} is the list of the data
$B_l, B_\natural, u_\natural$ and $\imath_l$ for each $\gh_n$.
We change the notation slightly from Section \ref{sec:intro} and \ref{sec:a22},
representing elements in $B = B_1$ with symbols inside a box or $\phi$.
For example the special (highest weight) element $1 \in B_1$
in the properties (I)--(II) is denoted by $\hako{1}$.
Consequently a state of the automata at each time is represented by
an element in
\begin{equation*}
{\mathcal P} = \{ b : {\mathbb Z} \rightarrow B \mid
b_j = \hako{1} \in B \text{ for } \vert j \vert \gg 1 \},
\end{equation*}
which is also interpreted as
\[
{\mathcal P} = \{ \cdots \otimes b_{-1}\otimes b_0 \otimes b_1 \otimes \cdots
\in \cdots \otimes B \otimes B \otimes B \otimes \cdots \mid
b_j = \hako{1} \text{ for } \vert j \vert \gg 1 \}.
\]
We let $T: {\mathcal P} \rightarrow {\mathcal P}$ denote the time
evolution operator as in Section \ref{subsec:cellauto}.
The essential data is the injection
\begin{equation*}
\imath_l : U'_q(\gh_{n-1})\text{-crystal }B_{l} \rightarrow
(U'_q(\gh_n)\text{-crystal }B_{1})^{\otimes l},
\end{equation*}
which will be utilized to label solitons
in the $U'_q(\gh_n)$ automata in terms of elements of
the $U'_q(\gh_{n-1})$-crystal $B_l$.

First we claim stable propagation of the 1-soliton as
\begin{theorem}\label{th:one}
For any $l \in \Z_{\ge 1}$ and $b \in U'_q(\gh_{n-1})\text{-crystal } B_l$,
the time evolution map $T$ acts on the state
\[
\cdots \otimes \hako{1} \otimes \hako{1} \otimes \imath_l(b) \otimes
\hako{1} \otimes \hako{1} \otimes \cdots
\]
as the overall translation to the right by $l$ slots.
\end{theorem}
Thus $l$ is the velocity of the soliton.
Even when $n$ is the minimal possible value for $\gh_n$,
Theorem \ref{th:one} makes sense if  $\imath_l$ and  $B_l$ for
`$U'_q(\gh_{n-1})$' are interpreted appropriately by an extrapolation of
the data given in Section \ref{subsec:data}.
For example, the solitons for $\gh_n = A^{(2)}_2$  case argued in
Section \ref{subsec:cellauto} agree with those
coming from `$U'_q(A^{(2)}_0)$-crystal $B_l$'  in such a sense.
The proof is given in Section \ref{subsec:1-solproof}.
For scattering of two solitons we have
\begin{theorem}\label{th:two}
Let $\gh_n = C^{(1)}_n \,(n \ge 3)$.
For fixed positive integers $l > k$, let
$B_l \ot B_k \ni  b_1 \otimes b_2 \simeq c_2 \ot c_1 \in B_k \ot B_l$  be an
isomorphism under the combinatorial $R$ matrix of the $U'_q(\gh_{n-1})$-crystals .
Then for $m \gg l$, there exists $t_0 > 0$ such that for any $t \ge t_0$ and some $m'$
\begin{eqnarray*}
T^t &:& \cdots \otimes \hako{1} \otimes \hako{1} \otimes
\imath_{l}(b_1) \otimes \hako{1}^{\,\otimes m} \otimes \imath_{k}(b_2)
\otimes \hako{1} \otimes \hako{1} \otimes \hako{1} \cdots \cd\cd\\
&\mapsto&
\cdots \cd \ot \hako{1} \otimes \hako{1} \otimes \hako{1} \otimes \imath_k(c_2) \otimes
\hako{1}^{\,\otimes m'} \otimes \imath_l(c_1) \otimes
\hako{1} \otimes \hako{1} \otimes \cdots.
\end{eqnarray*}
\end{theorem}
In other words, we have the combinatorial $R$ matrix as the
scattering matrix of the ultra-discrete solitons.
Compared with $\imath_l(b_1)$,
$\imath_k(c_2)$ is shifted to the right, but we do not concern the precise distance.
A sketch of a proof of Theorem \ref{th:two} will be given
in Section \ref{subsec:cproof}.

In fact we have a conjecture on $N$-soliton case
for general $\gh_n$.
\begin{conjecture}\label{con:multi}
Let $N \in {\mathbb Z}_{\ge 2}$.
Fix positive integers $k_1,\ldots, k_N$ ($k_1> \cd >k_N$) and the elements
$b_i \in B_{k_i}\,(i=1, \ldots, N)$ of the $U'_q(\gh_{n-1})$-crystals.
Define $c_i \in B_{k_i}$ by
$b_1 \ot \cd \ot b_N \simeq c_N \ot \cd \ot c_1$ under the
isomorphism
$B_{k_1} \ot B_{k_2} \ot \cd \ot B_{k_N} \simeq
B_{k_N} \ot \cd \ot B_{k_2} \ot B_{k_1}$.
For $m_1,\ldots, m_{N-1}  \gg k_1$,
there exists $t_0 > 0$ such that for any $t \ge t_0$,
{\small
\begin{eqnarray*}
T^t &:& \cdots \otimes \hako{1} \otimes
\imath_{k_1}(b_1) \otimes \hako{1}^{\,\otimes m_1} \otimes
\cd
\otimes \hako{1}^{\,\otimes m_{N-1}} \otimes
\imath_{k_N}(b_N) \otimes \hako{1}  \ot \cdots\\
&\mapsto&\quad
\cdots \otimes \hako{1} \otimes
\imath_{k_N}(c_N) \otimes \hako{1}^{\,\otimes m'_1} \otimes
\cd
\otimes \hako{1}^{\,\otimes m'_{N-1}} \otimes
\imath_{k_1}(c_1) \otimes \hako{1}  \ot \cdots,
\end{eqnarray*}
}
holds for some $m'_1, \ldots, m'_{N-1}$.
\end{conjecture}
In particular, $c_1, \ldots, c_N$
do not depend on $m_i$'s (i.e.,the order of collisions)
as long as $m_i \gg k_1$.
Again we do not concern the precise distance between $\imath_{k_1}(b_1)$
and $\imath_{k_N}(c_N)$ in the above.

The rank $n$ in Conjecture \ref{con:multi} should be taken greater than the
minimal possible values, i.e.,
$A^{(2)}_{2n-1}\; (n \ge 4), A^{(2)}_{2n}\; (n \ge 2),
B^{(1)}_n \;(n  \ge 4), C^{(1)}_n \;(n \ge 3), D^{(1)}_n\; (n \ge 5)$
and $D^{(2)}_{n+1}\; (n \ge 3)$.
We have checked the $N=2, 3$ cases by computer with several examples
for $\gh_n = A^{(2)}_7, A^{(2)}_4, A^{(2)}_6, B^{(1)}_4, D^{(1)}_5, D^{(2)}_4$
and the $N = 3$ case for $C^{(1)}_3$.

Let us present a few examples from the $U'_q(C^{(1)}_3)$ automaton.
We use the notation that will be introduced in Section \ref{subsec:data}.
The dynamical variables are taken from the crystal
$B_1 = \{\hako{1}, \hako{2}, \hako{3},
\hako{\bar{3}}, \hako{\bar{2}}, \hako{\bar{1}} \}$.
In the following examples we drop the boxes for simplicity.

\begin{example}\hfill
\begin{center}
t=0 : $\cdots 1  1 \bar{2} \bar{2} 3 2 1 1 1 1 1 \bar{3} 3 1 1 1 1 1 1 1 1 1 1 1 1 1 1 1 1 1 1 1 1 1\cdots $ \\
t=1 : $\cdots 1  1 1 1 1 1 \bar{2} \bar{2} 3 2 1 1 1 \bar{3} 3 1 1 1 1 1 1 1 1 1 1 1 1 1 1 1 1 1 1 1\cdots $ \\
t=2 : $\cdots 1  1 1 1 1 1 1 1 1 1 \bar{2} \bar{2} 3 2 1 \bar{3} 3 1 1 1 1 1 1 1 1 1 1 1 1 1 1 1 1 1\cdots $ \\
t=3 : $\cdots 1  1 1 1 1 1 1 1 1 1 1 1 1 1 \bar{2} 3 3 \bar{3} \bar{3} 3 1 1 1 1 1 1 1 1 1 1 1 1 1 1\cdots $ \\
t=4 : $\cdots 1  1 1 1 1 1 1 1 1 1 1 1 1 1 1 1 1 3 3 1 \bar{2} \bar{3} \bar{3} 3 1 1 1 1 1 1 1 1 1 1\cdots $ \\
t=5 : $\cdots 1  1 1 1 1 1 1 1 1 1 1 1 1 1 1 1 1 1 1 3 3 1 1 1 \bar{2} \bar{3} \bar{3} 3 1 1 1 1 1 1\cdots $ \\
t=6 : $\cdots 1  1 1 1 1 1 1 1 1 1 1 1 1 1 1 1 1 1 1 1 1 3 3 1 1 1 1 1 \bar{2} \bar{3} \bar{3} 3 1 1\cdots $
\end{center}
Compare this with
\begin{equation*}
(1,1,0,2) \ot (0,1,1,0) \simeq (0,2,0,0) \ot (0,1,2,1)
\end{equation*}
under the isomorphism  $B_4 \ot B_2 \simeq B_2 \ot B_4$
of $U'_q(C^{(1)}_2)$-crystals.
\end{example}

\begin{example}\hfill
\begin{center}
t=0 : $\cdots 1  1 \bar{1} \bar{2} \bar{2} 3 1 1 1 1 1 \bar{2} \bar{3} 2 1 1 1 1 1 1 1 1 1 1 1 1 1 1 1 1 1 1 1 1 1 1 1 1\cdots $ \\
t=1 : $\cdots 1  1 1 1 1 1 1 \bar{1} \bar{2} \bar{2} 3 1 1 1 \bar{2} \bar{3} 2 1 1 1 1 1 1 1 1 1 1 1 1 1 1 1 1 1 1 1 1 1\cdots $ \\
t=2 : $\cdots 1  1 1 1 1 1 1 1 1 1 1 1 \bar{1} \bar{2} 3 1 1 \bar{2} \bar{2} \bar{3} 2 1 1 1 1 1 1 1 1 1 1 1 1 1 1 1 1 1\cdots $ \\
t=3 : $\cdots 1  1 1 1 1 1 1 1 1 1 1 1 1 1 1 1 \bar{1} 3 1 1 1 \bar{2} \bar{2} \bar{2} \bar{3} 2 1 1 1 1 1 1 1 1 1 1 1 1\cdots $ \\
t=4 : $\cdots 1  1 1 1 1 1 1 1 1 1 1 1 1 1 1 1 1 1 1 \bar{1} 3 1 1 1 1 1 \bar{2} \bar{2} \bar{2} \bar{3} 2 1 1 1 1 1 1 1\cdots $ \\
t=5 : $\cdots 1  1 1 1 1 1 1 1 1 1 1 1 1 1 1 1 1 1 1 1 1 1 \bar{1} 3 1 1 1 1 1 1 1 \bar{2} \bar{2} \bar{2} \bar{3} 2 1 1\cdots $
\end{center}
Compare this with
\begin{equation*}
(0,1,0,2) \ot (1,0,1,1) \simeq (0,1,0,0) \ot (1,0,1,3)
\end{equation*}
under the  isomorphism $B_5 \ot B_3 \simeq B_3 \ot B_5$
of $U'_q(C^{(1)}_2)$-crystals.
\end{example}

\begin{example}\hfill
\begin{center}
t=0 : $\cdots 1  1 \bar{1} \bar{3} \bar{3} 3 3 1 1 1 1 1 \bar{1} \bar{2} 3 1 1 1 1 1 \bar{2} \bar{3} 1 1 1 1 1 1 1 1 1 1 1 1 1 1 1 1 1 1 1 1 1 1 1 1 1 1 1 1 1 1\cdots $ \\
t=1 : $\cdots 1  1 1 1 1 1 1 1 \bar{1} \bar{3} \bar{3} 3 3 1 1 1 \bar{1} \bar{2} 3 1 1 1 \bar{2} \bar{3} 1 1 1 1 1 1 1 1 1 1 1 1 1 1 1 1 1 1 1 1 1 1 1 1 1 1 1 1\cdots $ \\
t=2 : $\cdots 1  1 1 1 1 1 1 1 1 1 1 1 1 1 \bar{1} \bar{3} \bar{3} 3 1 1 \bar{1} \bar{2} 3 3 \bar{2} \bar{3} 1 1 1 1 1 1 1 1 1 1 1 1 1 1 1 1 1 1 1 1 1 1 1 1 1 1\cdots $ \\
t=3 : $\cdots 1  1 1 1 1 1 1 1 1 1 1 1 1 1 1 1 1 1 1 \bar{1} \bar{3} \bar{3} 1 1 3 3 1 \bar{1} \bar{2} \bar{2} \bar{2} 2 1 1 1 1 1 1 1 1 1 1 1 1 1 1 1 1 1 1 1 1\cdots $ \\
t=4 : $\cdots 1  1 1 1 1 1 1 1 1 1 1 1 1 1 1 1 1 1 1 1 1 1 1 \bar{1} 1 1 \bar{3} \bar{3} 3 3 1 1 1 \bar{1} \bar{2} \bar{2} \bar{2} 2 1 1 1 1 1 1 1 1 1 1 1 1 1 1\cdots $ \\
t=5 : $\cdots 1  1 1 1 1 1 1 1 1 1 1 1 1 1 1 1 1 1 1 1 1 1 1 1 1 \bar{1} 1 1 1 1 \bar{3} \bar{3} 3 3 1 1 1 1 1 \bar{1} \bar{2} \bar{2} \bar{2} 2 1 1 1 1 1 1 1 1\cdots $ \\
t=6 : $\cdots 1  1 1 1 1 1 1 1 1 1 1 1 1 1 1 1 1 1 1 1 1 1 1 1 1 1 1 \bar{1} 1 1 1 1 1 1 \bar{3} \bar{3} 3 3 1 1 1 1 1 1 1 \bar{1} \bar{2} \bar{2} \bar{2} 2 1 1\cdots $
\end{center}
Compare this with
\begin{equation*}
(0,2,2,0) \ot (0,1,0,1) \ot (0,0,1,1) \simeq
(0,0,0,0) \ot (0,2,2,0) \ot (1,0,0,3)
\end{equation*}
under the isomorphism
$B_6 \ot B_4 \ot B_2 \simeq B_2 \ot B_4 \ot B_6$
of $U'_q(C^{(1)}_2)$-crystals.
\end{example}

\subsection{Data on crystals}\label{subsec:data}
Let us present the data
$B_l, B_\natural, u_\natural$ and $\imath_l$ for each $\gh_n$.
$B_l$ and $B_\natural$ will be specified only as sets.
The crystal structure of $B_l$ is available in \cite{KKM}.
(See \cite{HKKOT} for $C^{(1)}_n$ case.)
For all the  $\gh_n$,  $u_\natural \in B_\natural$ is given by
\[
u_\natural = (\forall x_i, \bar{x}_i = 0)
\]
in the notation employed below.
As a set $B_l$ can be embedded into $B_\natural$ by
\begin{eqnarray*}
g_l : &B_l& \rightarrow B_\natural \\
&(x_i,\bar{x}_i)& \mapsto (x_i - l\delta_{1 i},\bar{x}_i).
\end{eqnarray*}
Obviously any $u \in B_\natural$ has the inverse image in each $B_l$
if $l$ is large enough.
It is easy to see that the composition
{\setlength{\tabcolsep}{2pt}
\begin{tabular}{ccccccc}
$B_\natural$ &
$\xrightarrow{g^{-1}_l}$ &
$B_l$ &
$\xrightarrow{\et_i, \ft_i}$ &
$B_l \sqcup \{0\}$ &
$\xrightarrow{g_l}$ &
$B_\natural\sqcup \{0\}$
\end{tabular}

\noindent
is independent of $l$ when $l$ is large enough.
Here we understand that $g_l(0) = 0$.
Thus one can endow $B_\natural$ with the actions of
$\et_i, \ft_i$ defined by this with $l$ sufficiently large.
(However it will not be used in this paper.)

%%%%%%%%%%%%%%%%%%%%%%%%%%%%%%%%%%%%%%%%%%%%%%%%%%%%%%%%%%%%
\noindent
$\gh_n=A^{(2)}_{2n-1}:$
\begin{eqnarray*} \label{eq:BofA2o}
B_l &=& \{ (x_{1},\dots ,x_{n},\bar{x}_{n},\dots ,\bar{x}_{1}) \in
\Z^{2n}  | x_{i}, \bar{x}_{i} \geq 0,\,
\sum_{i=1}^{n}(x_{i}+\bar{x}_{i}) =l \},\\
B_\natural &=& \{ (x_{1},\dots ,x_{n},\bar{x}_{n},\dots ,\bar{x}_{1}) \in
\Z^{2n}  | x_1 \leq 0,\, x_{i} \geq 0 \,(i \ne 1),\, \bar{x}_{i} \geq 0,\,
\sum_{i=1}^{n}(x_{i}+\bar{x}_{i}) =0 \}.
\end{eqnarray*}
For $B_1$ we use a simpler notation
\[
(x_{1},\dots ,x_{n},\bar{x}_{n},\dots ,\bar{x}_{1}) =
\begin{cases}
\hako{i} & \text{if }x_i=1,\,\text{others}=0, \\
\hako{\bar{i}} & \text{if }\bar{x}_i=1,\,\text{others}=0.
\end{cases}
\]
For $b=(x_1,\dots,x_{n-1},\bar{x}_{n-1},\dots,\bar{x}_1)$
in $U'_q(A^{(2)}_{2n-3})$-crystal $B_{l}$,
\begin{equation*}
\imath_l (b)=
	\hako{\bar{2}}^{\,\otimes \bar{x}_1} \otimes \dots \otimes
	\hako{\bar{n}}^{\,\otimes \bar{x}_{n\! -\! 1}} \otimes
	\hako{n}^{\,\otimes x_{n\! -\! 1}} \otimes \dots \otimes
	\hako{2}^{\,\otimes x_1}.
\end{equation*}
%%%%%%%%%%%%%%%%%%%%%%%%%%%%%%%%%%%%%%%%%%%%%%%%%%%%%%%%
\noindent
$\gh_n=A^{(2)}_{2n}:$
\begin{eqnarray*} \label{eq:BofA2e}
B_l &=& \{  (x_1,\dots,x_n,\bar{x}_n,\dots,\bar{x}_1) \in
\Z^{2n} \;| \; x_i,\bar{x}_i \ge 0,\, \sum_{i=1}^n (x_i + \bar{x}_i)
\le l \},\\
B_\natural &=& \{  (x_1,\dots,x_n,\bar{x}_n,\dots,\bar{x}_1) \in
\Z^{2n} \;| \; x_1 \leq 0,\, x_i \ge 0\,(i \ne 1),\,
\bar{x}_i \ge 0,\, \sum_{i=1}^n (x_i + \bar{x}_i) \le 0 \}.
\end{eqnarray*}
For $B_1$ we use a simpler notation
\[
(x_1,\dots,x_n,\bar{x}_n,\dots,\bar{x}_1) =
\begin{cases}
\hako{i} & \text{if }x_i=1,\,\text{others}=0, \\
\hako{\bar{i}} & \text{if }\bar{x}_i=1,\,\text{others}=0, \\
\phi & \text{if }x_i=0,\,\bar{x}_i=0 \text{ for all }i.
\end{cases}
\]
For $b=(x_1,\dots,x_{n-1},\bar{x}_{n-1},\dots,\bar{x}_1)$
in $U'_q(A^{(2)}_{2n-2})$-crystal $B_{l}$,
we define $s(b)=\sum_{i=1}^{n-1}(x_i+\bar{x}_i),\,
s'(b)=[ (l-s(b))/2 ]$.
\vspace*{10pt}\\
\noindent
If $l-s(b)$ is odd,
\[
\imath_l (b)=
	\phi \otimes \hako{\bar{1}}^{\,\otimes s'(b)} \otimes
	\hako{\bar{2}}^{\,\otimes \bar{x}_1} \otimes \dots \otimes
	\hako{\bar{n}}^{\,\otimes \bar{x}_{n\! -\! 1}} \otimes
	\hako{n}^{\,\otimes x_{n\! -\! 1}} \otimes \dots \otimes
	\hako{2}^{\,\otimes x_1} \otimes
	\hako{1}^{\,\otimes s'(b)},
\]
otherwise
\[
\imath_l (b)=
	\hako{\bar{1}}^{\,\otimes s'(b)} \otimes
	\hako{\bar{2}}^{\,\otimes \bar{x}_1} \otimes \dots \otimes
	\hako{\bar{n}}^{\,\otimes \bar{x}_{n\! -\! 1}} \otimes
	\hako{n}^{\,\otimes x_{n\! -\! 1}} \otimes \dots \otimes
	\hako{2}^{\,\otimes x_1} \otimes
	\hako{1}^{\,\otimes s'(b)}.
\]
When $n=1$, the above notation for $B_1$ and (\ref{eq:simpler})
are related by $1 = \hako{1}, 2 = \phi$ and $3 = \hako{\bar{1}}$.
The solitons and their velocity mentioned in the beginning of
Section \ref{subsec:cellauto} agree with the $n=1$ case here.
One interprets $s(b) = 0$ and $s'(b) = [l/2]$ for
$b$ from `$U'_q(A^{(2)}_0)$'-crystal $B_l$.
Then, under the identification $R = (1-(-1)^l)/2$ and $Q = R + s'(b)$,
the velocity is indeed $l = 2Q-R$.

%%%%%%%%%%%%%%%%%%%%%%%%%%%%%%%%%%%%%%%%%%%%%%%%%%%%%%%%%%%%%%%%%%%%%%%%%
\noindent
$\gh_n=B^{(1)}_{n}:$\\
\begin{eqnarray*} \label{eq:BofB}
B_l &=& \left\{ (x_{1},\dots ,x_{n},x_{0},\bar{x}_{n},\dots ,\bar{x}_{1}) \in
\Z^{2n} \times\{0,1\}
\left|
\begin{array}{c} x_{0} = 0\; \mbox{or} \; 1, x_{i}, \bar{x}_{i} \geq 0, \\
x_0 + \sum_{i=1}^{n}(x_{i}+\bar{x}_{i}) =l
\end{array}\right.
\right\},\\
B_\natural &=&
\left\{ (x_{1},\dots ,x_{n},x_{0},\bar{x}_{n},\dots ,\bar{x}_{1}) \in
\Z^{2n} \times\{0,1\} \left|
\begin{array}{c} x_{0} = 0\; \mbox{or} \; 1,\,x_1 \le 0,\\
x_{i} \ge 0\,(i \ne 1),\, \bar{x}_{i} \geq 0,\\
x_0+\sum_{i=1}^{n}(x_{i}+\bar{x}_{i}) =0
\end{array}\right.
\right\}.
\end{eqnarray*}
For $B_1$
we use a simpler notation
\[
(x_{1},\dots ,x_{n},x_{0},\bar{x}_{n},\dots ,\bar{x}_{1}) =
\begin{cases}
\hako{i} & \text{if }x_i=1,\,\text{others}=0, \\
\hako{\bar{i}} & \text{if }\bar{x}_i=1,\,\text{others}=0.
\end{cases}
\]
For $b=(x_1,\dots,x_{n-1},x_0,\bar{x}_{n-1},\dots,\bar{x}_1)$
in $U'_q(B^{(1)}_{n-1})$-crystal $B_{l}$,
\[
\imath_l (b)=
	\hako{\bar{2}}^{\,\otimes \bar{x}_1} \otimes \dots \otimes
	\hako{\bar{n}}^{\,\otimes \bar{x}_{n\! -\! 1}} \otimes
	\hako{0}^{\,\otimes x_0} \otimes
	\hako{n}^{\,\otimes x_{n\! -\! 1}} \otimes \dots \otimes
	\hako{2}^{\,\otimes x_1}.
\]
%%%%%%%%%%%%%%%%%%%%%%%%%%%%%%%%%%%%%%%%%%%%%%%%%%%%%%%%%%%%%%%%%%%%%%%%
\noindent
$\gh_n=C^{(1)}_{n}:$\\
\begin{eqnarray*} \label{eq:BofC}
B_l &=& \left\{  (x_1,\dots,x_n,\bar{x}_n,\dots,\bar{x}_1) \in
\Z^{2n} \Biggm|
\begin{array}{l}
x_i,\bar{x}_i \ge 0, \\
l \ge \sum_{i=1}^n
(x_i + \bar{x}_i) \in l - 2 \Z
\end{array}
\right\},\\
B_\natural &=& \left\{  (x_1,\dots,x_n,\bar{x}_n,\dots,\bar{x}_1) \in
\Z^{2n} \Biggm|
\begin{array}{l}
x_1 \le 0,\,x_i \ge 0 \, (i \ne 1), \, \bar{x}_i \ge 0, \\
0 \ge \sum_{i=1}^n
(x_i + \bar{x}_i) \in 2 \Z
\end{array}
\right\}.
\end{eqnarray*}
For $B_1$ we
use a simpler notation
\[
(x_1,\dots,x_n,\bar{x}_n,\dots,\bar{x}_1) =
\begin{cases}
\hako{i} & \text{if }x_i=1,\,\text{others}=0, \\
\hako{\bar{i}} & \text{if }\bar{x}_i=1,\,\text{others}=0.
\end{cases}
\]
For $b=(x_1,\dots,x_{n-1},\bar{x}_{n-1},\dots,\bar{x}_1)$
in $U'_q(C^{(1)}_{n-1})$-crystal $B_{l}$,
we define $s(b)=\sum_{i=1}^{n-1}(x_i+\bar{x}_i),\,
s'(b)=(l-s(b))/2$.
\[
\imath_l (b)=
	\hako{\bar{1}}^{\,\otimes s'(b)} \otimes
	\hako{\bar{2}}^{\,\otimes \bar{x}_1} \otimes \dots \otimes
	\hako{\bar{n}}^{\,\otimes \bar{x}_{n\! -\! 1}} \otimes
	\hako{n}^{\,\otimes x_{n\! -\! 1}} \otimes \dots \otimes
	\hako{2}^{\,\otimes x_1} \otimes
	\hako{1}^{\,\otimes s'(b)}.
\]
%%%%%%%%%%%%%%%%%%%%%%%%%%%%%%%%%%%%%%%%%%%%%%%%%%%%%%%%%%%%%%%%%%
\noindent
$\gh_n=D^{(1)}_{n}:$\\
\begin{eqnarray*} \label{eq:BofD}
B_l &=& \left\{ (x_{1},\dots ,x_{n},\bar{x}_{n},\dots ,\bar{x}_{1}) \in
\Z^{2n}  \left|
\begin{array}{c}
x_{n} = 0\; \mbox{or} \;\bar{x}_{n} =0,\; x_{i}, \bar{x}_{i} \geq 0,\\
\sum_{i=1}^{n}(x_{i}+\bar{x}_{i}) =l
\end{array}\right.
\right\},\\
B_\natural &=& \left\{ (x_{1},\dots ,x_{n},\bar{x}_{n},\dots ,\bar{x}_{1}) \in
\Z^{2n}  \left|
\begin{array}{c}
x_{n} = 0\; \mbox{or} \;\bar{x}_{n} =0,\; x_1 \le 0,\,\\
x_{i} \ge 0 \, (i \ne 1),\, \bar{x}_{i} \geq 0,\\
\sum_{i=1}^{n}(x_{i}+\bar{x}_{i}) = 0
\end{array}\right.
\right\}.
\end{eqnarray*}
For $B_1$
we use a simpler notation
\[
(x_{1},\dots ,x_{n},\bar{x}_{n},\dots ,\bar{x}_{1}) =
\begin{cases}
\hako{i} & \text{if }x_i=1,\,\text{others}=0, \\
\hako{\bar{i}} & \text{if }\bar{x}_i=1,\,\text{others}=0.
\end{cases}
\]
For $b=(x_1,\dots,x_{n-1},\bar{x}_{n-1},\dots,\bar{x}_1)$
in $U'_q(D^{(1)}_{n-1})$-crystal $B_{l}$,
\[
\imath_l (b)=
	\hako{\bar{2}}^{\,\otimes \bar{x}_1} \otimes \dots \otimes
	\hako{\bar{n}}^{\,\otimes \bar{x}_{n\! -\! 1}} \otimes
	\hako{n}^{\,\otimes x_{n\! -\! 1}} \otimes \dots \otimes
	\hako{2}^{\,\otimes x_1}.
\]
%%%%%%%%%%%%%%%%%%%%%%%%%%%%%%%%%%%%%%%%%%%%%%%%%%%%%%%%%%%%%%%%%%%%%%
\noindent
$\gh_n=D^{(2)}_{n+1}:$\\
\begin{eqnarray*} \label{eq:BofD2}
B_l &=& \left\{  (x_1,\dots,x_n,x_0,\bar{x}_n,\dots,\bar{x}_1) \in
\Z^{2n} \times \{0,1\} \Biggm|
\begin{array}{l}
x_0=\mbox{$0$ or $1$},x_i,\bar{x}_i \ge 0, \\
x_0 + \sum_{i=1}^n (x_i + \bar{x}_i) \le l
\end{array}
\right\},\\
B_\natural &=& \left\{  (x_1,\dots,x_n,x_0,\bar{x}_n,\dots,\bar{x}_1) \in
\Z^{2n} \times \{0,1\} \Biggm|
\begin{array}{l}
x_0=\mbox{$0$ or $1$},\,x_1 \le 0,\\
x_i \ge 0\, (i \ne 1),\, \bar{x}_i \ge 0, \\
x_0 + \sum_{i=1}^n (x_i + \bar{x}_i) \le 0
\end{array}
\right\}.
\end{eqnarray*}
For $B_1$
we use a simpler notation
\[
B_1 \ni (x_1,\dots,x_n,x_0,\bar{x}_n,\dots,\bar{x}_1) =
\begin{cases}
\hako{i} & \text{if }x_i=1,\,\text{others}=0, \\
\hako{\bar{i}} & \text{if }\bar{x}_i=1,\,\text{others}=0, \\
\phi & \text{if }x_i=0,\,\bar{x}_i=0 \text{ for all }i.
\end{cases}
\]
For $b=(x_1,\dots,x_{n-1},x_0,\bar{x}_{n-1},\dots,\bar{x}_1)$
in $U'_q(D^{(2)}_{n})$-crystal $B_{l}$,
we define $s(b)=\sum_{i=1}^{n-1}(x_i+\bar{x}_i),\,
s'(b)=[ (l-s(b))/2 ]$.
\vspace*{10pt}\\
\noindent
If $l-s(b)$ is odd,
\[
\imath_l (b)=
	\phi \otimes \hako{\bar{1}}^{\,\otimes s'(b)} \otimes
	\hako{\bar{2}}^{\,\otimes \bar{x}_1} \otimes \dots \otimes
	\hako{\bar{n}}^{\,\otimes \bar{x}_{n\! -\! 1}} \otimes
	\hako{0}^{\,\otimes x_0} \otimes
	\hako{n}^{\,\otimes x_{n\! -\! 1}} \otimes \dots \otimes
	\hako{2}^{\,\otimes x_1} \otimes
	\hako{1}^{\,\otimes s'(b)},
\]
otherwise
\[
\imath_l (b)=
	\hako{\bar{1}}^{\,\otimes s'(b)} \otimes
	\hako{\bar{2}}^{\,\otimes \bar{x}_1} \otimes \dots \otimes
	\hako{\bar{n}}^{\,\otimes \bar{x}_{n\! -\! 1}} \otimes
	\hako{0}^{\,\otimes x_0} \otimes
	\hako{n}^{\,\otimes x_{n\! -\! 1}} \otimes \dots \otimes
	\hako{2}^{\,\otimes x_1} \otimes
	\hako{1}^{\,\otimes s'(b)}.
\]

\subsection{Proof of Theorem \ref{th:one}}\label{subsec:1-solproof}
Our proof below uses the crystal structure of $B_l$ given
in \cite{HKKOT} for $\gh_n = C^{(1)}_n$ and in \cite{KKM} for the
other types.
\begin{lemma}\label{lem:i}
For $i \in \{1,2,\dots,n-1\}$, the following diagram is commutative:
\[
\begin{CD}
U'_q(\gh_{n-1})\text{-crystal }B_l @>{\imath_l}>>
U'_q(\gh_{n})\text{-crystal }B_1^{\otimes l}\\
@V{\tilde{e}_i}VV @VV{\tilde{e}_{i+1}}V \\
U'_q(\gh_{n-1})\text{-crystal }B_l \sqcup\{0\} @>>{\imath_l}>
U'_q(\gh_{n})\text{-crystal }B_1^{\otimes l}\sqcup\{0\}.\\
\end{CD}
\]
The same relation holds between $\ft_i$ and $\ft_{i+1}$.
\end{lemma}
The Kashiwara operators of
$U'_q(\gh_{n-1})$ and $U'_q(\gh_n)$-crystals
should not be confused although we use the same notation.
The proof of the lemma is due to the explicit
rules for $\tilde{e}_i$ \cite{KKM, HKKOT} and
the embedding of $U'_q(\gh_{n-1})\text{-crystal }B_l$ into
$U'_q(\gh_{n-1})\text{-crystal }B_1^{\ot l}$ as
$U_q(\g_{n-1})$-crystals \cite{KN}.
Here $\g_{n-1}$ is the classical subalgebra of $\gh_{n-1}$.
According to $\gh_n = A^{(2)}_{2n-1}, A^{(2)}_{2n},
B^{(1)}_n, C^{(1)}_n, D^{(1)}_n$ and $D^{(2)}_{n+1}$
it is given by
$\geh_n = C_n, C_n, B_n , C_n, D_n$ and $B_n$ respectively.
\vskip0.2cm
\noindent
{\em Proof of Theorem \ref{th:one}}.\\
$\gh_n=A^{(2)}_{2n-1}$:\\
Since $B_l$ of $U'_q(A^{(2)}_{2n-3})$-crystal
is isomorphic to $B(l \Lambda_1)$ as
$U_q(C_{n-1})$-crystals, for any $b \in B_l$ there exists a sequence
$i_1,\dots,i_p  \in \{1,\dots,n-1\}$ such that
$b_0=\tilde{e}_{i_p} \cdots \tilde{e}_{i_1} b$  with
$b_0=(l,0,\dots,0) \in B_l$.
{}From Lemma \ref{lem:i},
$\imath_l(b_0)=\tilde{e}_{i_p+1} \cdots \tilde{e}_{i_1+1} \imath_l(b)$
is valid.

Taking sufficiently large $L$ and ${\mathfrak f}=
\tilde{f}^l_2 \cdots \tilde{f}^l_{n-1} \tilde{f}^l_{n}
\tilde{f}^l_{n-1} \cdots \tilde{f}^l_{2}$, one has
\[
\begin{CD}
u_L \otimes \imath_l(b) \otimes \hako{1}^{\otimes l} @.
\hako{1}^{\otimes l} \ot \imath_l(b) \otimes u_L
\\
@V{\tilde{e}_{i_p+1} \cd \tilde{e}_{i_1+1}}VV
@V{\tilde{e}_{i_p+1} \cd \tilde{e}_{i_1+1}}VV
\\
u_L \otimes \imath_l(b_0) \otimes \hako{1}^{\otimes l} @.
\hako{1}^{\otimes l} \ot \imath_l(b_0) \otimes u_L
\\
@| @| \\
(L,0,\dots,0) \otimes \hako{2}^{\otimes l} \otimes \hako{1}^{\otimes l} @.
\hako{1}^{\ot l} \ot \hako{2}^{\ot l} \ot (L,0,\dots,0)
\\
@V{\mathfrak f}VV   @V{\mathfrak f}VV
\\
(L,0,\dots,0) \otimes \hako{\bar{2}}^{\ot l} \ot \hako{1}^{\ot l} @.
\hako{1}^{\ot l} \ot \hako{\bar{2}}^{\ot l} \ot (L,0,\dots,0)
\\
@V{\tilde{e}_0^{L} \tilde{f}_1^{L+2l}}VV
@V{\tilde{e}_0^{L} \tilde{f}_1^{L+2l}}VV
\\
(0,\dots,0,L) \otimes \hako{\bar{1}}^{\ot l} \ot \hako{2}^{\ot l} @.
\hako{\bar{1}}^{\ot 2l} \ot (0,l,0,\dots,0,L-l)\\
@V{\mathfrak f}VV  @V{\mathfrak f}VV
\\
(0,\dots,0,L) \otimes \hako{\bar{1}}^{\ot l} \ot \hako{\bar{2}}^{\ot l}@.
\hako{\bar{1}}^{\ot 2l} \ot (0,\dots,0,l,L-l)
\\
@V{\tilde{e}_1^{L+l} \tilde{f}_0^{L+2l}}VV
@V{\tilde{e}_1^{L+l} \tilde{f}_0^{L+2l}}VV
\\
(L,0,\dots,0) \otimes \hako{1}^{\ot 2l}
@>{\sim}>>
\hako{1}^{\ot 2l} \ot (L,0,\dots,0).
\\
\end{CD}
\]
Thus the isomorphism $B_L \ot B_1^{\ot 2l}
\overset{\sim}{\rightarrow} B_1^{\ot 2l} \ot B_L$
sends $u_L \otimes \imath_l(b) \otimes \hako{1}^{\otimes l}$ to
$\hako{1}^{\otimes l} \ot \imath_l(b) \otimes u_L$.
This proves Theorem \ref{th:one}.

\noindent
$\gh_n=A^{(2)}_{2n}$:\\
Let $b^{(m)}_0=(m,0,\dots,0) \in B_l$ of $U'_q(A_{2n-2}^{(2)})$,
where $0 \le m \le l$.
First we consider $b = b^{(l)}_0$ case.
Taking sufficiently large $L$ and ${\mathfrak f}=
\tilde{f}^l_2 \cd \tilde{f}^l_{n-1} \tilde{f}^l_{n}
\tilde{f}^l_{n-1} \cd \tilde{f}^l_{2}$, one has
\[
\begin{CD}
u_L \otimes \imath_l(b^{(l)}_0) \otimes \hako{1}^{\otimes l} @.
\hako{1}^{\otimes l} \ot \imath_l(b^{(l)}_0) \otimes u_L
\\
@| @| \\
(L,0,\dots,0) \otimes \hako{2}^{\otimes l} \otimes \hako{1}^{\otimes l} @.
\hako{1}^{\ot l} \ot \hako{2}^{\ot l} \ot (L,0,\dots,0)
\\
@V{\mathfrak f}VV   @V{\mathfrak f}VV
\\
(L,0,\dots,0) \otimes \hako{\bar{2}}^{\ot l} \ot \hako{1}^{\ot l} @.
\hako{1}^{\ot l} \ot \hako{\bar{2}}^{\ot l} \ot (L,0,\dots,0)
\\
@V{\tilde{e}_{1}^{L} \tilde{f}_0^{2l} \tilde{f}_1^{L+2l}}VV
@V{\tilde{e}_{1}^{L} \tilde{f}_0^{2l} \tilde{f}_1^{L+2l}}VV
\\
(L,0,\dots,0) \otimes \hako{1}^{\ot l} \ot \hako{2}^{\ot l} @.
\hako{1}^{\ot 2l} \ot (L-l,l,0,\dots,0)\\
@V{\mathfrak f}VV  @V{\mathfrak f}VV
\\
(L,0,\dots,0) \otimes \hako{1}^{\ot l} \ot \hako{\bar{2}}^{\ot l}@.
\hako{1}^{\ot 2l} \ot (L-l,0,\dots,0,l,0)
\\
@V{\tilde{e}_{1}^{L+l} \tilde{f}_{0}^{2l} \tilde{f}_{1}^{L+2l}}VV
@V{\tilde{e}_{1}^{L+l} \tilde{f}_{0}^{2l} \tilde{f}_{1}^{L+2l}}VV
\\
(L,0,\dots,0) \otimes \hako{1}^{\ot 2l}
@>{\sim}>>
\hako{1}^{\ot 2l} \ot (L,0,\dots,0).
\\
\end{CD}
\]
Thus the isomorphism $B_L \ot B_1^{\ot 2l}
\overset{\sim}{\rightarrow} B_1^{\ot 2l} \ot B_L$ sends
$u_L \otimes \imath_l(b^{(l)}_0) \otimes \hako{1}^{\otimes l}$ to
$\hako{1}^{\otimes l} \ot \imath_l(b^{(l)}_0) \otimes u_L$, verifying
Theorem \ref{th:one} for $b = b^{(l)}_0$.
Explicitly it reads
\begin{equation}\label{eq:2toright}
(L,0, \ldots, 0) \ot \hako{2}^{\ot l} \ot \hako{1}^{\ot l}
\simeq \hako{1}^{\ot l} \ot \hako{2}^{\ot l}\ot(L,0,\ldots,0).
\end{equation}

Let us proceed to arbitrary $b \in B_l$ case.
We will attribute the proof  to the $b = b^{(m)}_0$ case
for some $0 \le m \le l$ and
eventually to the $b^{(l)}_0$ case shown above.
Set $k=[\frac{l-m}{2}], k'=[\frac{l-m+1}{2}]$ so that
$\imath_l(b^{(m)}_0) = \phi^{\ot k'-k}
\ot \hako{\bar{1}}^{\ot k}\ot \hako{2}^{\ot m} \ot \hako{1}^{\ot k}$
according to Section \ref{subsec:data}.
Since $B_l$ of $U'_q(A^{(2)}_{2n-2})$-crystal
is isomorphic to
$B(l \Lambda_1) \oplus B((l-1) \Lambda_1) \oplus
\cd \oplus B(0)$ as
$U_q(C_{n-1})$-crystals, for any $b \in B_l$ there exists a sequence
$i_1,\dots,i_p \in \{1,\dots,n-1\}$ such that
$b^{(m)}_0=\tilde{e}_{i_p} \cd \tilde{e}_{i_1} b$ for
some $0 \le m \le l$.
{}From Lemma \ref{lem:i} one has
\[
\begin{CD}
u_L \otimes \imath_l(b) \otimes \hako{1}^{\otimes l+k'} @.
\hako{1}^{\otimes l} \ot \imath_l(b) \ot \hako{1}^{\otimes k'} \ot u_L
\\
@V{\tilde{e}_{i_p+1} \cd \tilde{e}_{i_1+1}}VV
@V{\tilde{e}_{i_p+1} \cd \tilde{e}_{i_1+1}}VV
\\
u_L \otimes \imath_l(b^{(m)}_0) \otimes \hako{1}^{\otimes l+k'} @.
\hako{1}^{\otimes l} \ot \imath_l(b^{(m)}_0) \ot \hako{1}^{\otimes k'} \ot u_L
\\
@V{\tilde{e}_{1}^{L+k'} \tilde{f}_{0}^{k+k'} \tilde{f}_{1}^{L+k+2k'}}VV
@V{\tilde{e}_{1}^{L+k'} \tilde{f}_{0}^{k+k'} \tilde{f}_{1}^{L+k+2k'}}VV
\\
(L,0,\dots,0) \ot \hako{1}^{\ot k'} \ot \hako{2}^{\ot l} \ot \hako{1}^{\ot l}
@>{\sim}>>
\hako{1}^{\ot k'} \ot \hako{1}^{\ot l} \ot \hako{2}^{\ot l} \ot (L,0,\dots,0),
\end{CD}
\]
where in the bottom we have used (\ref{eq:2toright}) and
$u_L \ot \hako{1} \simeq \hako{1} \ot u_L$.
Thus the isomorphism $B_L \ot B_1^{\ot 2l} \tilde{\rightarrow}
B_1^{\ot 2l} \ot B_L$
sends $u_L \otimes \imath_l(b) \otimes \hako{1}^{\otimes l}$ to
$\hako{1}^{\otimes l} \ot \imath_l(b) \otimes u_L$, proving
Theorem \ref{th:one}.

\noindent
$\gh_n=B^{(1)}_{n}$:\\
The proof is similar to the case $\gh_n=A^{(2)}_{2n-1}$
with ${\mathfrak f}=
\tilde{f}^l_2 \cd \tilde{f}^l_{n-1} \tilde{f}^{2l}_{n}
\tilde{f}^l_{n-1} \cd \tilde{f}^l_{2}$.

\noindent
$\gh_n=C^{(1)}_{n}$:\\
The proof is similar to  the case $\gh_n=A^{(2)}_{2n}$.

\noindent
$\gh_n=D^{(1)}_{n}$:\\
The proof is similar to the case $\gh_n=A^{(2)}_{2n-1}$
with ${\mathfrak f}=
\tilde{f}^l_2 \cd \tilde{f}^l_{n-2} \tilde{f}^{l}_{n}
\tilde{f}^l_{n-1} \cd \tilde{f}^l_{2}$.

\noindent
$\gh_n=D^{(2)}_{n+1}$:\\
The proof is similar to the case  $\gh_n=A^{(2)}_{2n}$
with ${\mathfrak f}=
\tilde{f}^l_2 \cd \tilde{f}^l_{n-1} \tilde{f}^{2l}_{n}
\tilde{f}^l_{n-1} \cd \tilde{f}^l_{2}$.
\begin{flushright}
$\square$
\end{flushright}

\subsection{Proof of Theorem \ref{th:two}}\label{subsec:cproof}

We divide the proof into Part I and Part II.
The statements in Part I are valid not only for $C^{(1)}_n$ but for any $\gh_n$.
Apart from the separation into two solitons in the final state,
this already proves Theorem \ref{th:two} for those $\gh_n$ in which
the decomposition of $B_l \ot B_k$ is multiplicity-free.
Among the list of $\gh_n$ in question, such cases are
$A^{(1)}_{2n-1}, B^{(1)}_n$ and $D^{(1)}_n$.
Since the  $C^{(1)}_n$ case  has the multiplicity,
we need Part II, which relies on the explicit
result on the combinatorial $R$ matrices for $U'_q(C^{(1)}_n)$ in \cite{HKOT}.
In the rest of Section \ref{sec:conj}
we shall write $U'_q(\gh_n)$-crystals as $B_l$  and
$U'_q(\gh_{n-1})$-crystals as $\tilde{B}_l$ for distinction.

\vskip0.2cm
{\em Part I}.
For sufficiently large $L$, let $u_L = (L, 0, \ldots, 0) \in B_L$ be the
highest weight element.
Let ${\mathcal R}: \tilde{B}_l \ot \tilde{B}_k \simeq
\tilde{B}_k \ot \tilde{B}_l$
be the combinatorial $R$ matrix of
$U'_q(\gh_{n-1})$-crystals.
The main claim in Part I is
\begin{proposition}\label{pr:split}
For
$b_1 \in \tilde{B}_l, b_2 \in \tilde{B}_k\, (l > k)$
and  $L_1, L_2 \in {\mathbb Z}_{\ge 1}$,
suppose there exist $t_0 \in {\mathbb Z}_{\ge 1},\, c_1 \in
\tilde{B}_l$ and $c_2 \in \tilde{B}_k$ such that
\begin{align}
&u^{\ot t}_L \ot \fbox{$1$}^{\ot L_1} \ot \imath_l(b_1) \ot
\fbox{$1$}^{\ot L_2}
 \ot \imath_k(b_2) \ot \fbox{$1$}^{\ot L_3}\nonumber\\
&\quad\;\,\simeq
\fbox{$1$}^{\ot L'_1} \ot \imath_k(c_2) \ot \fbox{$1$}^{\ot L'_2}
 \ot \imath_l(c_1) \ot \fbox{$1$}^{\ot L'_3}\ot u^{\ot t}_L \label{eq:sendu}
\end{align}
holds for any $t \ge t_0$ for some $L'_1,\, L'_2, \, L_3$ and $L'_3$
under the isomorphism
$B_L^{\ot t} \ot B_1 \ot \cd \ot B_1 \simeq B_1 \ot \cd \ot B_1 \ot B_L^{\ot t}$.
Define ${\mathcal S}: \tilde{B}_l \ot \tilde{B}_k
\rightarrow \tilde{B}_k \ot \tilde{B}_l$ through
this relation by ${\mathcal S}(b_1\ot b_2) = c_2 \ot c_1$.
Then ${\mathcal R}(b_1 \ot b_2) = {\mathcal S}(b_1 \ot b_2)$
implies  ${\mathcal R}(\ft_i(b_1 \ot b_2)) = {\mathcal S}(\ft_i(b_1 \ot b_2))$
for each $i = 1, \ldots, n-1$.
Similarly, $\et_i(b_1 \ot b_2) = 0$ implies
$\et_i{\mathcal S}(b_1 \ot b_2) = 0$ for each $i = 1, \ldots, n-1$.
\end{proposition}
\begin{lemma}\label{cor:cor1}
For each $i = 1, \ldots, n-1$, we have a commutative diagram:
\begin{equation*}
\begin{CD}
\tilde{B}_l \ot \tilde{B}_k @>{\imath_l \ot \imath_k}>> (B_1)^{\ot l+k}\\
@V{\tilde{e}_i}VV @VV{\et_{i+1}}V\\
(\tilde{B}_l \ot \tilde{B}_k)
\sqcup\{0\}@>{\imath_l \ot \imath_k}>> (B_1)^{\ot l+k}\sqcup\{0\}.
\end{CD}
\end{equation*}
The same relation holds also between  $\ft_i$ and $\ft_{i+1}$.
\end{lemma}
This is a simple corollary of Lemma \ref{lem:i}.
Similarly we have
\begin{lemma}\label{lem:lem2}
Let $a_1,\ldots, a_L \in B_1,\,$
$i_1,\dots,i_m$ be the subsequence of $1,\dots,L$
satisfying $a_j = \fbox{$1$} \in B_1$ if $j \not\in \{i_1,\dots,i_m\}$
$(m \le L)$,
and $p_k=a_{i_k}$ $(k=1,\dots,m)$.
%Let $p_1,\ldots, p_m
%\in B_1$ be the subsequence of
%$a_1, \ldots, a_L \in B_1$ consisting of all the elements that are not
%$\fbox{$1$} \in B_1$ $(m \le L)$.
%%
%Assume the same relation between
%$p'_1, \ldots, p'_m \in B_1$ and
%$a'_1, \ldots, a'_L \in B_1$.
%
Then for any $t, t' \in {\mathbb Z}_{\ge 0}$ and
$L \in {\mathbb Z}_{\ge 1}$,
%the two relations
%\begin{align*}
%&\ft_{i+1}(p_1 \ot \cd \ot p_m ) = p'_1 \ot \cd \ot p'_m,\\
%&\ft_{i+1}(u_L^{\ot t} \ot a_1 \ot \cd \ot a_L \ot u_L^{\ot t'}) =
%u_L^{\ot t} \ot a'_1 \ot \cd \ot a'_L \ot u_L^{\ot t'}
%\end{align*}
%are equivalent
%for each $i = 1, \ldots, n-1$,
%where $p'_k=a'_{i_k}$.
\[
\begin{array}{cl}
1.&%\text{if }
\ft_{i+1}(p_1 \ot \cd \ot p_m ) = p'_1 \ot \cd \ot p'_m\\
%\text{ ,then }\\
&\Rightarrow
\ft_{i+1}(u_L^{\ot t} \ot a_1 \ot \cd \ot a_L \ot u_L^{\ot t'}) =
u_L^{\ot t} \ot a'_1 \ot \cd \ot a'_L \ot u_L^{\ot t'},\\
& \hspace{1cm} \text{where }
a'_j=
\begin{cases}
p'_k & \text{ if } j=i_k,\\
\fbox{$1$} & \text{otherwise},
\end{cases}\\[15pt]
2.&%\text{if }
\ft_{i+1}(u_L^{\ot t} \ot a_1 \ot \cd \ot a_L \ot u_L^{\ot t'}) =
u_L^{\ot t} \ot a'_1 \ot \cd \ot a'_L \ot u_L^{\ot t'}
%\text{ ,then }
\\
&\Rightarrow
\ft_{i+1}(p_1 \ot \cd \ot p_m ) = p'_1 \ot \cd \ot p'_m,\;\;
\text{where }p'_k=a'_{i_k},\\[15pt]
3.&
\ft_{i+1}(u_L^{\ot t} \ot a_1 \ot \cd \ot a_L \ot u_L^{\ot t'}) = 0\;
\Leftrightarrow\;
\ft_{i+1}(p_1 \ot \cd \ot p_m )=0,
\end{array}
\]
for each $i = 1, \ldots, n-1$.
The same is true also for $\et_{i+1}$.
\end{lemma}

{\em Proof of Proposition \ref{pr:split}}.
Suppose ${\mathcal R}(b_1 \ot b_2 ) = {\mathcal S}(b_1 \ot b_2) = c_2 \ot c_1$.
Thus $(c_2, c_1) \in \tilde{B}_k \times \tilde{B}_l$ is related to $(b_1, b_2)$ via
(\ref{eq:sendu}).
Set $b'_1 \ot b'_2 = \ft_i(b_1 \ot b_2)$ for some $i = 1, \ldots, n-1$
and put $c'_2 \ot c'_1 = {\mathcal S}(b'_1 \ot b'_2)$.
We are to show ${\mathcal R}(b'_1 \ot b'_2) = c'_2 \ot c'_1$.
{}From Lemma \ref{cor:cor1} one has
\begin{equation}
\ft_{i+1}\left(\imath_l(b_1) \ot \imath_k(b_2)\right) =
\imath_l(b'_1) \ot \imath_k(b'_2). \label{eq:fb}
\end{equation}
Setting
$t'=0$,
$a_1 \ot \cd \ot a_L =
\fbox{1}^{\ot L_1} \ot \imath_l(b_1) \ot \fbox{1}^{\ot L_2}
 \ot \imath_k(b_2) \ot \fbox{1}^{\ot L_3}$ and
$p_1 \ot \cd \ot p_m = \imath_l(b_1)\ot \imath_k(b_2)$
in Lemma \ref{lem:lem2}, one has
\begin{align}
&\ft_{i+1}\left(u^{\ot t}_L \ot \fbox{1}^{\ot L_1}
\ot \imath_l(b_1) \ot \fbox{1}^{\ot L_2}
 \ot \imath_k(b_2) \ot \fbox{1}^{\ot L_3}\right) \nonumber\\
&\quad\;\, = u^{\ot t}_L \ot \fbox{1}^{\ot L_1}
\ot \imath_l(b'_1) \ot \fbox{1}^{\ot L_2}
 \ot \imath_k(b'_2) \ot \fbox{1}^{\ot L_3}, \nonumber
\end{align}
where (\ref{eq:fb}) is used.
By sending $u_L$ to the right as in
(\ref{eq:sendu}), this is equivalent to
\begin{align}
&\ft_{i+1}\left(\fbox{1}^{\ot L'_1} \ot \imath_k(c_2) \ot \fbox{1}^{\ot L'_2}
 \ot \imath_l(c_1) \ot \fbox{1}^{\ot L'_3}\ot u^{\ot t}_L\right) \nonumber\\
&\quad\;\, = \fbox{1}^{\ot L'_1} \ot \imath_k(c'_2) \ot \fbox{1}^{\ot L'_2}
 \ot \imath_l(c'_1) \ot \fbox{1}^{\ot L'_3}\ot u^{\ot t}_L \nonumber
\end{align}
in terms of the $(c'_2, c'_1)$ specified above.
With the help of Lemma \ref{lem:lem2} and  \ref{cor:cor1}
we can now go backwards to see that this implies
$\ft_{i+1}(\imath_k(c_2) \ot \imath_l(c_1)) =
\imath_k(c'_2) \ot \imath_l(c'_1)$ hence
$\ft_i(c_2 \ot c_1) = c'_2 \ot c'_1$.
Thus we have
${\mathcal R}(b'_1 \ot b'_2) = {\mathcal R}(\ft_i(b_1 \ot b_2)) =
\ft_i {\mathcal R}(b_1 \ot b_2) = \ft_i(c_2 \ot c_1) = c'_2 \ot c'_1$.
It is very similar to verify that
$\et_i(b_1 \ot b_2) = 0$ implies $\et_i{\mathcal S}(b_1 \ot b_2) = 0$
for any  $i = 1, \ldots, n-1$.
\hfill
$\square$

\vskip.5cm
\noindent
The above proof is also valid in $A^{(1)}_n$ case \cite{HHIKTT}.
Viewed as $U_q(\geh_{n-1})$-crystals with Kashiwara operators
$\ft_i, \et_i \, (1 \le i \le n-1)$, both crystals
$\tilde{B}_l \ot \tilde{B}_k$ and  $\tilde{B}_k \ot \tilde{B}_l$
decompose into connected components.
Note that (\ref{eq:sendu}) obviously tells that $U_q(\geh_{n-1})$-weights of
$b_1 \ot b_2$ and ${\mathcal S}(b_1\ot b_2)=c_2 \ot c_1$ are equal.
Therefore apart from the
separation into two solitons in the final state, Proposition \ref{pr:split}
reduces the proof of Theorem \ref{th:two} to showing
${\mathcal R}(b_1 \ot b_2) = {\mathcal S}(b_1 \ot b_2)$ only
for the $U_q(\geh_{n-1})$-highest weight elements $b_1\ot b_2$.
In particular, if the tensor product decomposition
of $\tilde{B}_l \ot \tilde{B}_k$ is multiplicity-free,
it only remains to check the separation.

\vskip0.2cm
{\em Part II}. Here we concentrate on the $U'_q(C^{(1)}_n)$ automaton and
prove the separation and
${\mathcal R}(b_1 \ot b_2) = {\mathcal S}(b_1 \ot b_2)$
directly for $U_q(C_{n-1})$-highest weight elements $b_1 \ot b_2$ ($n \ge 3$).
In the notation
$(x_1, \ldots, x_{n-1},\bar{x}_{n-1}, \ldots, \bar{x}_1) \in \tilde{B}_k$ or
$\tilde{B}_l$, they have the form
\begin{equation}\label{eq:hwe}
(f,0,\ldots,0) \ot (d,c,0,\ldots,0,b) \in \tilde{B}_l \ot \tilde{B}_k,
\quad f  \ge b + c.
\end{equation}
In what follows we always assume $l > k$ and use the non-negative integers
$e$ and $a$ defined by $l = f+2e$ and $k= 2a+b+c+d$.
\begin{proposition}[\cite{HKOT}]\label{pr:hwe}
Under the isomorphism ${\mathcal R}: \tilde{B}_l\ot \tilde{B}_k \simeq
\tilde{B}_k \ot \tilde{B}_l$ of $U'_q(C^{(1)}_{n-1})$-crystals,
the image of (\ref{eq:hwe}) is given by
\begin{equation}\label{eq:seikai}
(k-2e,0,\ldots,0) \ot (d+l-k-y,c,0,\ldots, 0,b-y),\quad
y = \min(l-k,(b-d)_+),
\end{equation}
if $a \ge e$.
In case  $a < e$, it is given by
\begin{align*}
&(k-2a,0,\ldots,0) \ot (d+l-k-e+a-z,c,0,\ldots, 0,b+e-a-z),\\
&z = \min(b-d+e-a,l-k-e+a),\\
\intertext{if $l-k> e-a \ge d-b$, and}
&(k-2e+2w,0,\ldots,0) \ot (d+l-k-w,c,0,\ldots, 0,b+w),\\
&w = \min(l-k,(2e-2a-d+b)_+),
\end{align*}
otherwise.
\end{proposition}
See Section \ref{subsec:a22crystal} for the definition of the
symbol $(x)_+$.
Below we employ  the notation
\begin{equation*}
[s,t,u] = (x_1 = L-s-t-2u, x_2 = s, \bar{x}_1 = t, \mbox{ other } x_i, \bar{x}_i = 0)
\in B_L,
\end{equation*}
and always assume $L \gg s, t, u$.
To save the space, the tensor product of $U'_q(C^{(1)}_n)$-crystals
such as
\begin{equation*}
[s,t,u]^{\ot j} \ot \hako{\bar{1}}^{\ot p} \ot \hako{2}^{\ot q}
\ot [s'',t'',u''] \ot \hako{3}^{\ot r}\ot
[s',t',u']^{\ot j'}
\end{equation*}
will simply be denoted by
\begin{equation*}
[s,t,u]^j\, \bar{1}^p \,2^q \,[s'',t'',u''] \,3^r\,[s',t',u']^{j'}.
\end{equation*}
{}From the results in \cite{HKOT} we further derive two lemmas given below.
\begin{lemma}\label{lem:ikko}
If $L \gg p, q, s, t, u$, we have
\begin{align*}
[0,0,0] \,\bar{1}^p \,2^q &\simeq 1^{p+q}\,[q,0,p],\\
[s,t,u]\, 1 &\simeq \begin{cases}
1 \, [s,t+1,u-1] & u > 0, \\
\bar{1}\, [s,t-1,0] & u = 0, t > 0,\\
2 \,[s-1,0,0] &u = t = 0, s > 0,\\
1\, [0,0,0] &u = t = s = 0.
\end{cases}
\end{align*}
\end{lemma}
\begin{lemma}\label{lem:aiueo}
If $L \gg s, t, u, a, b, c, d$, we have
\begin{alignat*}{3}
&[s,t,u] \,\bar{1}^a \,\bar{2}^b \,3^c \,2^d \, 1^\delta &&\\
&\simeq
\bar{1}^{a}\, 2^{s+t-a} \,1^{u-t-s+2a+b+c+d} \,
\bar{1}^{a+u} \,\bar{2}^{t-a+b} \,3^c \, 2^{d}\,[0,0,0]
\quad  &&\mbox{if } (\mathrm{I}), \\
&\simeq
\bar{1}^{a}\, 2^{t-a+2b+c}\,1^{u-t+2a-b+d} \,
\bar{1}^{a+u} \,\bar{2}^{t-a+b} \,3^c \,2^{s-2b-c+d}\,[0,0,0]
\quad  &&\mbox{if } (\mathrm{II}),\\
&\simeq
\bar{1}^{a}\, 2^{b+c+d}\, 1^{a+u} \,
\bar{1}^{s+t+u-b-c-d}\, \bar{2}^{2b+c+d-s} \,3^c\, 2^{d}\,[0,0,0]
\quad  &&\mbox{if } (\mathrm{III}), \\
&\simeq
\bar{1}^{a} \,2^{b+c+d} \,1^{a+u} \,
\bar{1}^{t+u+b-d}\, \bar{2}^{d}\, 3^c \,2^{s-2b-c+d}\,[0,0,0]
\quad  &&\mbox{if } (\mathrm{IV}), \\
&\simeq
\bar{1}^{t} \,2^{s}\, 1^{u-s-t+2a+b+c+d} \,
\bar{1}^{a+u} \,\bar{2}^{b} \,3^c \,2^{d}\,[0,0,0]
 \quad  &&\mbox{if } (\mathrm{V}),\\
&\simeq
\bar{1}^{t} \,2^{a+2b+c-t} \,1^{u+a-b+d} \,
\bar{1}^{a+u} \,\bar{2}^{b} \,3^c \,2^{s+t-a-2b-c+d}\,[0,0,0]
\quad  &&\mbox{if } (\mathrm{VI}), \\
&\simeq
\bar{1}^{t} \,2^{a+b+c+d-t} \,1^{a+u}\,
\bar{1}^{u+s+t-b-c-d}\, \bar{2}^{a+2b+c+d-s-t} \,3^c \,2^{d}\,[0,0,0]
\quad  &&\mbox{if } (\mathrm{VII}),\\
&\simeq
\bar{1}^{t}\, 2^{a+b+c+d-t} \,1^{a+u} \,
\bar{1}^{u+a+b-d}\, \bar{2}^{d} \,3^c \,2^{s+t-a-2b-c+d}\,[0,0,0]
\quad  &&\mbox{if } (\mathrm{VIII}),
\end{alignat*}
where the conditions $(\mathrm{I})$--$(\mathrm{VIII})$ are given by
\begin{align*}
(\mathrm{I}):& \;t \ge a,\; s \le 2b+c,\; s+t \le a + b + c + d,\\
(\mathrm{II}):& \;t \ge a,\; s > 2b+c,\; t \le a-b+d,\\
(\mathrm{III}):& \;t \ge a,\; s \le 2b+c,\; s+t > a+b+c+d,\\
(\mathrm{IV}):& \;t \ge a,\; s > 2b+c,\; t > a-b+d,\\
(\mathrm{V}):& \;t < a,\; s+t \le \min(a+2b+c,a+b+c+d),\\
(\mathrm{VI}):& \;t < a,\; b \le d,\; s+t > a+2b+c,\\
(\mathrm{VII}):& \;t < a,\; b > d,\; a+b+c+d < s+t \le a+2b+c,\\
(\mathrm{VIII}):& \;t < a,\; b > d,\; s+t > a+2b+c.
\end{align*}
On the boundaries, the formulae in the adjacent domains are equal.
The domain (II) is absent if $b > d$.
$\delta = a+b+c+d+t+2u$ for $(\mathrm{I,III})$,
$a-b+d+s+t+2u$ for $(\mathrm{II,IV,VI,VIII})$,
$2a+b+c+d+2u$ for $(\mathrm{V,VII})$.
\end{lemma}
{\em Proof of Theorem \ref{th:two}.}
Let $b_1 \ot b_2$ be the highest weight element (\ref{eq:hwe}) and
$c_2 \ot c_1 = {\mathcal R}(b_1 \ot b_2)$ be its image specified in
Proposition \ref{pr:hwe}.
Two solitons labeled by $b_1$ and $b_2$ emerge in our automaton
as the patterns $\imath_l(b_1) = \bar{1}^e \,2^f\,1^e$ and
$\imath_k(b_2) = \bar{1}^a \bar{2}^b 3^c 2^d 1^a$, respectively.
Thus  ${\mathcal S}(b_1\ot b_2)$ can be found by computing the image of
\begin{equation}\label{eq:start}
[0,0,0]^j\, \bar{1}^e \, 2^f \, 1^{e+m} \, \bar{1}^a \,
\bar{2}^b \, 3^c \, 2^d \, 11\cdots
\end{equation}
under the isomorphism
$B^{\ot j}_L \ot (B_1\ot B_1 \ot \cd )
 \simeq (B_1 \ot B_1 \ot \cd )\ot B^{\ot j}_L$
for an interval $m$.
This calculation can be done
by using Lemma \ref{lem:ikko} and \ref{lem:aiueo} only.
It branches into numerous sectors depending on the seven  parameters
$a, b, c, d, f, e$ and $m$.
(We use the dependent variables $l = f+2e$ and $k = 2a+b+c+d$ simultaneously.)
We have checked case by case that the collision always `completes' within
two time steps ($j=2$) ending with the result
$1^{m_1} \imath_k(c_2) 1^{m_2} \imath_l(c_1) 11\ldots  [0,0,0]^j$ for some
$m_1$ and $m_2$.
This establishes ${\mathcal R} = {\mathcal S}$
hence Theorem \ref{th:two}.
Below we illustrate a typical such calculation
in the sector
\begin{equation}\label{eq:ineq}
a \ge 2e, \quad b > d, \quad
a-e \le m  \le f + e - a -2b - c.
\end{equation}
In this case
$c_2 \ot c_1 = {\mathcal R}(b_1 \ot b_2)$ is determined by
Proposition \ref{pr:hwe} as
\begin{equation*}
c_1 = (l-k-b+2d,c,0,\ldots,0,d) \in \tilde{B}_l,
\quad
c_2 = (k-2e,0,\ldots, 0) \in \tilde{B}_k.
\end{equation*}
Thus we seek the patterns
\begin{equation}\label{eq:finalsoliton}
\imath_k(c_2) = \bar{1}^e \,2^{k-2e} \,1^e, \quad
\imath_l(c_1) = \bar{1}^{a+b-d}\, \bar{2}^d \,3^c \,2^{l-k-b+2d} \,1^{a+b-d}
\end{equation}
to show up  after a collision and separate from each other.
Starting from (\ref{eq:start}) with $j=3$,
one can rewrite it by means of  Lemma \ref{lem:ikko} and (\ref{eq:ineq}) as
\begin{equation*}
[0,0,0]^2\, 1^{\delta_1} \bar{1}^e 2^{m-e} [s_1,0,0]
\bar{1}^a \bar{2}^b 3^c 2^d 11\ldots,
\end{equation*}
with $s_1 = f+e-m$ and $\delta_1 = f+2e$.
{}From (\ref{eq:ineq}) one can apply  Lemma \ref{lem:aiueo} (VIII)
to transform this into
\begin{equation*}
[0,0,0]^2 1^{\delta_1} \bar{1}^e 2^{s'_2} 1^a \,
\bar{1}^{a_2} \bar{2}^d 3^c 2^{d_2} 11\ldots  [0,0,0],
\end{equation*}
where $s'_2 = m-e+a+b+c+d, a_2 = a+b-d$ and
$d_2 = s_1-a-2b-c+d$.
This completes the calculation of the one time step in the automaton.
The second time step can be worked out  similarly.
Lemma \ref{lem:ikko} with (\ref{eq:ineq})
transforms the above into
\begin{equation*}
[0,0,0] 1^{\delta_2} \bar{1}^e 2^{a-2e} [s_2,0,0]\,
 \bar{1}^{a_2} \bar{2}^d 3^c 2^{d_2}\,11\ldots [0,0,0],
\end{equation*}
where $s_2 = s'_2 - a + 2e = m+b+c+d+e$ and
$\delta_2 = \delta_1 + 2e + s'_2$.
This time Lemma \ref{lem:aiueo} (VI) is applied, leading to
\begin{equation}\label{eq:after}
[0,0,0] 1^{\delta_2} \bar{1}^e 2^{k-2e} 1^{\delta_3}\,
 \bar{1}^{a+b-d} \bar{2}^d 3^c 2^{l-k-b+2d}\,11\ldots [0,0,0]^2,
\end{equation}
where $\delta_3 = a_2-d+d_2 = f+e-m-b-c-d$.
This already contains the patterns (\ref{eq:finalsoliton}).
Let us calculate one more time step to confirm that they are
separating hereafter.
Since $\delta_3 > 2e$, (\ref{eq:after}) is isomorphic to
\begin{equation}\label{eq:after2}
1^{\delta_4} \bar{1}^e [k-2e,0,0] 1^{\delta_3 - 2e}
\bar{1}^{a+b-d} \bar{2}^d 3^c 2^{l-k-b+2d}\,11\ldots [0,0,0]^2,
\end{equation}
where $\delta_4 = \delta_2 + k$.
If $k \le \delta_3$ this becomes
\begin{equation}\label{eq:after3}
1^{\delta_4} \bar{1}^e 2^{k-2e} 1^{\delta_3+l-k }\,
 \bar{1}^{a+b-d} \bar{2}^d 3^c 2^{l-k-b+2d}\,11\ldots [0,0,0]^3.
\end{equation}
Compared with (\ref{eq:after}), the distance of the two patterns here
has increased by $l-k$ in accordance with the velocity
of solitons in Theorem \ref{th:one}.
In the other case $k > \delta_3$ one rewrites (\ref{eq:after2}) as
\begin{equation*}
1^{\delta_4} \bar{1}^e 2^{\delta_3-2e} [k-\delta_3,0,0]
\bar{1}^{a+b-d} \bar{2}^d 3^c 2^{l-k-b+2d}\,11\ldots [0,0,0]^2,
\end{equation*}
for which Lemma \ref{lem:aiueo} (V) can be applied.
The result coincides with (\ref{eq:after3}).
\begin{flushright}
$\square$
\end{flushright}
%

%%%%%%%%%%%%%%%%%%%%%%%%%%%%%%%%%%%%%%%%%%%%
\section{Discussion}\label{sec:dis}

In this paper we have proposed a class of
cellular automata associated with $U_q'(\gh_n)$.
They are essentially solvable
vertex models at $q=0$ in the vicinity of the
ferromagnetic vacuum.
They exhibit soliton behavior
stated in Theorem \ref{th:one}, \ref{th:two} and Conjecture \ref{con:multi}.

Behavior of tensor products of crystals in the vicinity
of ferromagnetic vacuum has not been investigated in detail so far.
To understand it better will be a key
to prove Conjecture \ref{con:multi} and
to clarify
the relevance to the subalgebra $\gh_{n-1}$,
which has  also been observed in the RSOS models in
the ferromagnetic-like `regime II' \cite{BR,K}.

The automata considered in this paper are associated with
an $l \rightarrow \infty$  limit
$R': B_\natural \ot B_1 \rightarrow B_1 \ot B_\natural$ of
the ordinary combinatorial $R$ matrices $R: B_l \ot B_1 \simeq B_1 \ot B_l$.
To vary $B_1$ from site to site and to replace $R'$ with
the corresponding combinatorial $R$ matrices
is a natural generalization.
See \cite{HHIKTT}.

We close by raising a few questions.
What kind of soliton equations can possibly be related to the
$U'_q(\gh_n)$ automata in general?
Is it possible to derive them  conceptually from the associated vertex models,
bypassing the route;
vertex models $\xrightarrow{q \rightarrow 0}$ automata
$\xrightarrow{\text{tropical variable change}}$ soliton equations?
Is it efficient to seek a piecewise analytic form of the combinatorial
$R$ matrices $B_l \ot B_k \simeq B_k \ot B_l$
by ultra-discretizing the relevant solitons solutions?
Is it possible to extract informations on the associated energy function
from the automata?
%
%%%%%%%%%%%%%%%%%%%%%%%%%%%%%%%%%%%%%%%%%%%
\appendix

\section{The
map $R': B_\natural \otimes B_1 \rightarrow  B_1 \otimes B_\natural$}
\label{app:natural}

Let $g_l: B_l \rightarrow B_\natural$ be the embedding of $B_l$ as a set as
mentioned in Section \ref{sec:conj}.
By the definition, for any $u \in B_\natural$
there exists the
unique element $u_l \in B_l$ such that
$g_l(u_l) = u$ for each $l$ which is sufficiently large.
Consider the composition:

{\setlength{\tabcolsep}{2pt}
\begin{tabular}{ccccccc}
$B_\natural \times B_1$ &
$\xrightarrow{g^{-1}_l\otimes id}$ &
$B_l \ot B_1$ &
$\xrightarrow{\et_i, \ft_i}$ &
$(B_l \ot B_1) \sqcup \{0\}$ &
$\xrightarrow{g_l \times id}$ &
$(B_\natural \times B_1)\sqcup \{0\}$ \\
$(u,  b)$ &
$\longmapsto$ &
$u_l \ot b$ &
$\longmapsto$ &
$u'_l\ot b'$ &
$\longmapsto$ &
$(g_l(u_l'), b')$,
\end{tabular}

\noindent
where of course either $u'_l = u_l$ or $b' = b$, and we interpret
$(g_l\times id)(0)  = 0$.
For all the $B_l$ considered in this paper it is easy to see that
this is independent of $l$
if  $l$ is sufficiently large.
We let $B_\natural \ot B_1$ denote the set
$B_\natural \times B_1$ equipped with the actions
$\tilde{e}'_i, \tilde{f}'_i$  determined as above with
sufficiently large $l$.
We shall simply write it as $\tilde{e}'_i, \tilde{f}'_i
: B_\natural \ot B_1 \rightarrow
(B_\natural \ot B_1) \sqcup \{0\}$.
Similarly we define $B_1 \ot B_\natural$ and
$\tilde{e}'_i, \tilde{f}'_i: B_1 \ot B_\natural \rightarrow
(B_1 \ot B_\natural) \sqcup \{0\}$.

Given $u \in B_\natural$ and $b \in B_1$, let $c = c(l,u,b) \in B_1$ and
$v= v(l,u,b) \in B_l$ be the elements determined by
\begin{equation*}
u_l \ot b \simeq c \ot v
\end{equation*}
under the combinatorial $R$ matrix
$R: B_l \ot B_1 \simeq B_1 \ot B_l$.
Then we have
\begin{conjecture}\label{con:atability}
For any fixed $u \in B_\natural$ and $b \in B_1$,
the elements $c(l,u,b) \in B_1$ and
$g_l\bigl(v(l,u,b)\bigr) \in B_\natural$ are independent of $l$
for $l$  sufficiently large.
\end{conjecture}
Assuming the conjecture we define the map $R':
 B_\natural \otimes B_1 \rightarrow  B_1 \otimes B_\natural$ to be
the composition:

{\setlength{\tabcolsep}{2pt}
\begin{tabular}{cccccccc}
$R'$ &:
$B_\natural \ot B_1$ &
$\xrightarrow{g^{-1}_l\ot id}$ &
$B_l \ot B_1$ &
$\overset{R}{\simeq}$ &
$B_1 \ot B_l$ &
$\xrightarrow{id \ot g_l}$ &
$B_1 \ot B_\natural$ \\
&
$u \ot b$ &
$\longmapsto$ &
$u_l \ot b$ &
$\simeq$ &
$c \ot v$ &
$\longmapsto$ &
$c \ot g_l(v)$,
\end{tabular}

\noindent
with  $l$ sufficiently large.
Obviously this is invertible.
Moreover it commutes with $\tilde{e}'_i$ and $\tilde{f}'_i$,
although this fact is not used in the main text.
For $\tilde{e}'_i$ this can be seen from the commutative diagram
($\tilde{f}'_i$ case is completely parallel.)
\begin{equation*}
\begin{CD}
B_\natural \ot B_1 @>{g_l^{-1}\ot id}>> B_l \ot B_1 @>{R}>>
B_1 \ot B_l @>{id \ot g_l}>> B \ot B_\natural\\
@V{\tilde{e}'_i}VV @V{\et_i}VV @VV{\et_i}V @VV{\tilde{e}'_i}V\\
B_\natural \ot B_1 @>{g_l^{-1}\ot id}>>
B_l \ot B_1 @>{R}>>
B_1 \ot B_l @>{id \ot g_l}>>
B \ot B_\natural,
\end{CD}
\end{equation*}
where the left and the right squares are the definitions of
$\tilde{e}'_i$ (for elements not annihilated by $\tilde{e}'_i$).
The map $R'$ indeed has the property (I) in
Section \ref{sec:intro}.
The property (II) is also valid for $A^{(2)}_2$ and $C^{(1)}_n$.
We conjecture it  for all the
cases considered in this paper.
%%%%%%%%%%%%%%%%%%%%%%%%%%%%%%%%%%%%%%%%%%%%%
\vskip0.4cm
\noindent
{\bf Note}:  While writing the paper the authors learned from \cite{FOY} that
the energy of combinatorial $R$ matrices for $U'_q(A^{(1)}_n)$
is encoded in the phase shift of  soliton scattering in the automaton.
We thank Yasuhiko Yamada for communicating their result.

%%%%%%%%%%%%%%%%%%%%%%%%%%%%%%%%%%%%%%%%%%%%

\end{document}